\documentclass[11pt] {article}
\usepackage[a4paper,margin=1in]{geometry}
\usepackage{lineno}

\usepackage[T1]{fontenc} %
\usepackage[normalem]{ulem} %

\usepackage[usenames,dvipsnames,svgnames,x11names]{xcolor}

\usepackage{cite}

\usepackage{balance}

\usepackage{listings}

\lstdefinelanguage{XML}
{
basicstyle=\ttfamily\footnotesize,
  morestring=[b]",
  moredelim=[s][\bfseries\color{Maroon}]{<}{\ },
  moredelim=[s][\bfseries\color{Maroon}]{</}{>},
  moredelim=[l][\bfseries\color{Maroon}]{/>},
  moredelim=[l][\bfseries\color{Maroon}]{>},
  morecomment=[s]{<?}{?>},
  morecomment=[s]{<!--}{-->},
  commentstyle=\color{gray},
  stringstyle=\color{blue},
  identifierstyle=\color{red}
}

\usepackage{moreverb}

\usepackage[nounderscore]{syntax}

\usepackage[pdftex]{graphicx}
\graphicspath{{./figures/}}
\DeclareGraphicsExtensions{.pdf}

\usepackage[cmex10]{amsmath}
\usepackage{amssymb}
\usepackage{mathtools}
\usepackage{amsthm}
\usepackage{amsfonts}
\usepackage{gensymb}

\usepackage{subfig} %

\usepackage{algorithmicx}
\usepackage{algpseudocode}
\usepackage[ruled]{algorithm}
\definecolor{light-gray}{gray}{0.75}
\algrenewcommand{\algorithmiccomment}[1]{\hskip3em{{\footnotesize \textcolor{light-gray}{$\blacktriangleright$}}} #1}

\usepackage{multirow} %
\usepackage{rotating} %
\usepackage{booktabs} %
\usepackage{colortbl} %
\usepackage{tablefootnote} %

\usepackage{array}
\newcolumntype{L}[1]{>{\raggedright\let\newline\\\arraybackslash\hspace{0pt}}m{#1}}
\newcolumntype{C}[1]{>{\centering\let\newline\\\arraybackslash\hspace{0pt}}m{#1}}
\newcolumntype{R}[1]{>{\raggedleft\let\newline\\\arraybackslash\hspace{0pt}}m{#1}}

\usepackage[pdftex,colorlinks=true,urlcolor=blue,citecolor=blue]{hyperref}

\usepackage{xspace}

\usepackage{enumitem}

\newcommand{\at}{\textsc{ATLAS}\xspace}

\usepackage{blindtext}

\def\orcid#1{\kern .08em\href{https://orcid.org/#1}{\includegraphics[keepaspectratio,width=0.7em]{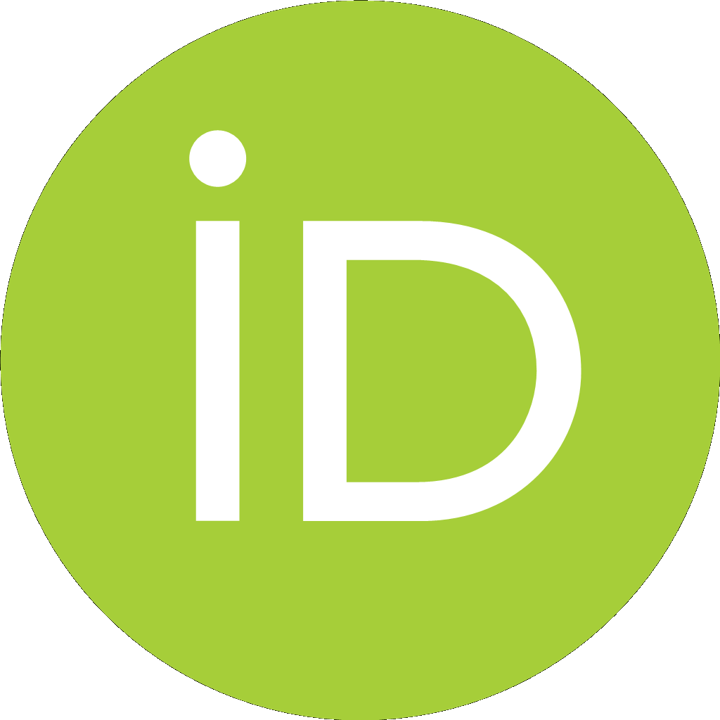}}}

\begin{document}
\title{\at: Efficient Out-of-Core Inference for Billion-Scale Graph Neural Networks~\thanks{~Preprint of paper to appear in the proceedings of The 35th International Symposium on High-Performance Parallel and Distributed Computing
(HPDC 26): Pranjal Naman and Yogesh Simmhan, “\at: Efficient Out-of-Core Inference for Billion-Scale Graph Neural Networks”, in \textit{International Symposium on High-Performance Parallel and Distributed Computing (HPDC)}, 2026. DOI: \url{https://doi.org/10.1145/3806645.3807597}}}

\author{Pranjal Naman\orcid{0009-0000-9912-9522} and Yogesh Simmhan$^1$\orcid{0000-0003-4140-7774}\\~\\
\em Department of Computational and Data Sciences (CDS),\\
\em Indian Institute of Science (IISc),\\
\em Bangalore 560012 India\\~\\
\texttt{Email:\{pranjalnaman, simmhan\}@iisc.ac.in}
}
\date{}
\maketitle

\begin{abstract}
  Graph Neural Network (GNN) inference on billion-scale graphs is critical for domains like fintech and recommendation systems.
  Full-graph inference on these large graphs can be challenging due to high communication costs in distributed settings and high I/O costs in disk-backed Out-of-Core~(OOC) settings. 
  Existing OOC systems, operating across disk and memory, primarily focus on GNN training and perform poorly for full-graph inference due to massive read amplification, irregular I/O and memory pressure.
  We present \at, a disk-based GNN inference framework that enables efficient full-graph, layer-wise inference on graphs whose topologies, features and intermediate embeddings exceed the available memory on single machines.
  \at replaces \textit{gather-based} execution with a \textit{broadcast-based} model that enables sequential, single-pass streaming reads of features and embeddings per layer.
  A tiered memory--disk hierarchy with minimum-pending-message eviction, graph reordering and a GPU-accelerated pipeline sustains high throughput within $128$\,GiB RAM and $2$\,TiB SSD.
  Across out-of-core graphs with up to $4$B edges and $550$\,GiB features and multiple GNN architectures, 
  \at improves end-to-end inference time by $\approx12$--$30\times$ over State-of-the-Art (SOTA) OOC baselines on a single workstation, while remaining within $\approx5\%$
  when features fit in memory.
\end{abstract}

\section{Introduction}\label{sec:intro}

Graph neural networks~(GNNs) have become popular for learning low-dimensional representations from linked data, capturing both the topology and associated features~\cite{kipf2017gcn, hamilton2017sage}. This makes them especially adept at performing a wide range of tasks such as detecting financial fraud in transaction networks~\cite{liu2021pick, dou2020enhancing,bharadwaj2026npci}, predicting traffic flows and signaling in Intelligent Transportation Systems (ITS)~\cite{guo2019attention,simmhan2026scale}, and
making e-commerce recommendations~\cite{yang2021consisrec}.
GNN inference typically follows a \textit{message-passing} paradigm, where each vertex \textit{gathers} and \textit{aggregates} messages from its $k$-hop neighborhood (\textit{computation graph}) and applies a neural-network transformation per layer.
This recursive multi-hop aggregation makes inference (and training) expensive due to irregular graph memory access and per-layer neural network compute.

\paragraph*{Motivation}
Given these memory and computational expenses, optimizing inference is crucial for real-world deployments. GNNs are often deployed in settings where the underlying graph can change, or the GNN model may be updated, requiring predictions to be refreshed periodically.
Even in systems that support incremental inference~\cite{ripple, ripple++, inkstream}, a full inference pass over the entire~(materialized historical) graph is still required for a new GNN model to establish a baseline against which incremental updates can be applied to future graph entities. While this process is not latency-critical, it must complete within a reasonable time, e.g., a couple of hours rather than days.

This is further complicated by the fact that GNNs often find applications in real-world \textit{large-scale} graphs comprising millions to billions of vertices and edges, e.g., for predicting fraud in a fintech transaction graph or recommending products or friends in e-commerce or social networks.
While the graph topology may fit in the RAM of a single machine, the feature and embedding vectors associated with vertices and edges are large and often exceed the available memory on workstations and even servers. E.g., the \textit{IGB-Large} citation graph~\cite{igb} used in our experiments has $100$M vertices, $1.2$B edges, and $1024$-length FP16 features, which together \textit{consume $219$\,GiB of RAM}~(Table~\ref{tab:datasets}) and well-exceed the RAM~($128$\,GiB) and GPU memory~($32$\,GiB) on our RTX 5090 GPU workstation.

To circumvent this, several works adopt a \textit{distributed data-parallel approach}, where multiple compute servers collaboratively train or infer a GNN model on smaller, distributed subgraphs~\cite{dsp, distdgl, p3, aligraph, namanEuropar, naman2026optimes}. 
Meanwhile, distributed GNN inference systems~\cite{kaler2023communication,chen2025deal} 
employ strategies like probabilistic caching of frequently accessed remote features, and collaborative graph and feature tensor partitioning.

\textit{Disk-based GNN training and inference} has emerged as a promising alternative~\cite{diskgnn, caliex, capsule, outre, ginex}, along the lines of prior \textit{Out-of-Core~(OOC)} disk-based parallel graph processing~\cite{vora2019lumos,xu2020hybrid,roy2013x}.
These focus on efficient data layouts and intelligent caching strategies to fully utilize memory, disk capacity and bandwidth on a single machine. This is all the more relevant given the lower latency and higher bandwidth of Solid State Disks~(SSDs) with capacities of 2\,TiB+ common even for prosumer disks, at a much lower price point than RAM.

\paragraph*{Challenges}
While disk-based OOC GNN training has been widely studied recently, efficient OOC inference poses a variety of challenges, as discussed below, and remains fairly underexplored. 
\begin{enumerate}[leftmargin=0pt,itemindent=20pt,label={\em(\arabic*)},topsep=0pt,listparindent=\parindent,] 
    \item \textit{Working-set Amplification During Inference.}
    Out-of-core GNN training methods optimize multiple aspects of the training pipeline, including \textit{data transfer}, such as leveraging GPU-direct~\cite{hyperion, gids} to bypass the CPU in the I/O path; \textit{data organization}~\cite{diskgnn, ginex, capsule, outre}, which governs the ordering of data movement and the use of multi-tier caching; and the \textit{data storage layout}~\cite{diskgnn}, where the logical arrangement of features is redesigned to enable more efficient training. While inference, at first glance, just appears to be the forward pass phase of training, these are fundamentally different workloads.
    
    Training typically operates on a small subset of vertices which have labels available~(e.g., $\approx1\%$ for \textit{OGBN-Papers100M}~\cite{ogb} in Table~\ref{tab:datasets}), whereas inference computes representations for the \textit{entire vertex set}. This changes the access pattern from repeated sampling of localized neighborhoods to materializing embeddings for all vertices. Existing OOC training frameworks exploit the small working set during training to perform optimizations, often requiring the entire computation graph to be precomputed~\cite{diskgnn, capsule}. Applying this strategy to inference is impractical, as constructing full unsampled computation graphs for the entire graph upfront is \textit{prohibitively expensive} in time and storage, e.g., on \textit{OGBN-Papers100M} for a 3-layer GNN using DGL~\cite{wang2019deep} with disk-backed topology takes $\approx1$\,hour and $\approx500$\,GiB of storage for just $1\%$ of vertices.

\item \textit{Sampling Effects on Inference.} During training, GNNs commonly use \textit{neighborhood sampling}~\cite{hamilton2017sage} at each layer to limit memory usage by considering only a subset of neighbors. However, such sampling is undesirable during inference, as it leads to non-deterministic outputs~\cite{inferturbo, kaler2022accelerating, ripple}, which is unacceptable for accuracy-critical applications such as fintech and ITS. As a result, inference typically requires \textit{full-graph} execution, where each vertex aggregates information from its complete neighborhood.
In disk-resident settings, this results in repeated random accesses to vertex features and intermediate embeddings for \textit{all} vertices, rather than a comparatively smaller working set for training, e.g., in \textit{IGB-Large}~\cite{igb}, training a 2-layer GNN on $1\%$ training vertices with neighborhood sampling touches only $\approx8\%$ of all vertices per epoch~\cite{naman2025gpu}.
When the feature set fits in RAM, layer-wise baselines can be competitive due to caching, but performance degrades sharply once features exceed memory and disk I/O dominates.

\item \textit{I/O Challenges at Full-Graph Scale.}
A common approach to GNN inference is to reuse the training pipeline without the backward pass, often referred to as \textit{vertex-wise inference}. In this setting, each vertex~(or \textit{batch} of vertices) recursively gathers features from its $k$-hop neighbors to compute its embedding. However, this approach is highly inefficient for GNN inference~(Fig.~\ref{fig:example}).
First, it leads to \textit{random accesses} to vertex features. Since features are stored on disk and accessed at block granularity~(e.g., 4\,KB), even small, scattered reads fetch entire blocks, reducing effective bandwidth. 
Second, it incurs \textit{repeated accesses}, where the same vertex features are fetched multiple times across different target vertices~(or batches). E.g., a high-degree vertex may be read repeatedly when processing each of its neighbors, further amplifying disk I/O.
Third, it performs \textit{redundant computation}, as overlapping neighborhoods are recomputed across batches.
Some works mitigate these issues using graph partitioning~\cite{marius, distdgl}, where subgraphs are processed in memory to eliminate disk accesses. However, this results in dropping cross-partition dependencies, which impacts model accuracy.

An alternative is \textit{layer-wise inference}, where embeddings for all vertices are computed \textit{one layer at a time}. This eliminates redundant computation by sharing intermediate results across vertices. However, random and repeated accesses still exist, since each vertex continues to independently gather features from its in-neighbors within a layer.
For instance, DGI~\cite{dgi}, a layer-wise inference framework with SSD-backed execution, still performs random feature and embedding gathers, leading to significant I/O overhead. Despite operating in a layer-wise manner, it does not address the fundamental issue of repeated and irregular data access.

We observe that the root cause across both vertex-wise and layer-wise approaches is their reliance on a \textit{gather-based} execution model. Since each vertex independently pulls data from its neighbors, the system remains input-unaware, leading to redundant data movement, poor locality, and severe I/O amplification.
\end{enumerate}

\begin{figure}[t]
    \centering
    \subfloat[PA/SAGE2/5090\label{subfig:motivation-1}]{
        \includegraphics[width=0.28\columnwidth]{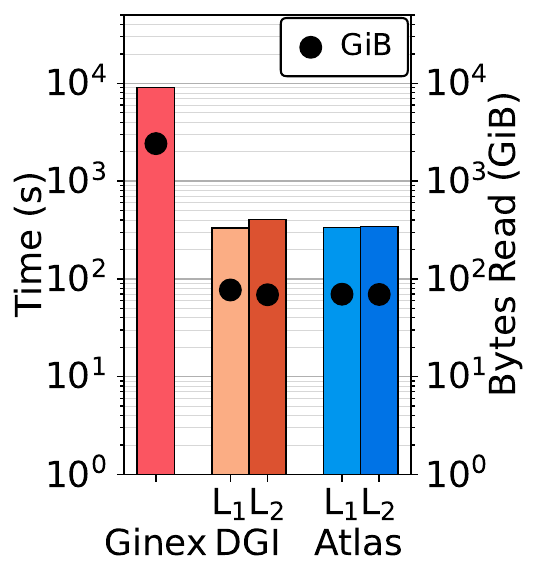}
    }\hfill
    \subfloat[MA/SAGE2/5090\label{subfig:motivation-2}]{
        \includegraphics[width=0.28\columnwidth]{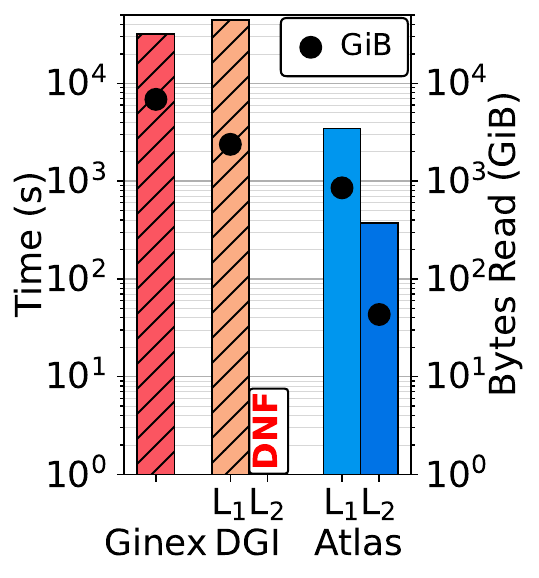}
    }\hfill
    \subfloat[PA/SAGE2/4090\label{subfig:motivation-3}]{
        \includegraphics[width=0.24\columnwidth]{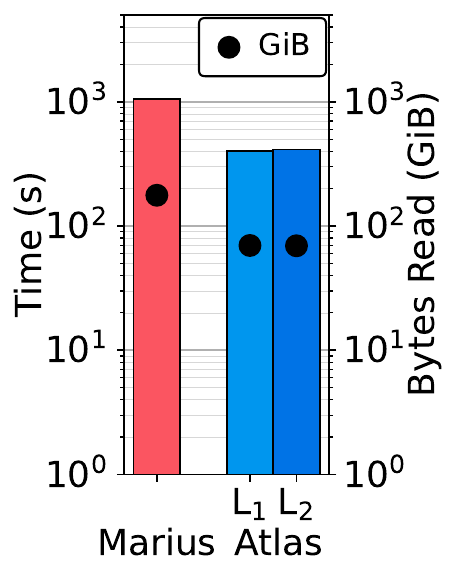}
    }
    \caption{Time taken~(left Y axis, \textit{hatched bars}: extrapolated, \textit{solid bars}: completed) and Total bytes read from disk~(\textit{markers}, right Y axis), for full-graph inference of 2-layer \textit{SAGEConv}, with topology and features disk-resident for Ginex~\cite{ginex}, DGI~\cite{dgi}, Marius~\cite{marius} and \at~(ours), for \textit{\underline{Pa}pers}~\cite{ogb} and \textit{\underline{MA}G240M-Cites}~\cite{ogb}, on 4090 and 5090 GPU workstations.}
    \label{fig:motivation}
\end{figure}

We \textit{empirically demonstrate} these challenges in Fig.~\ref{fig:motivation} by comparing three SOTA systems: \textit{DGI}~\cite{dgi}~(layer-wise inference), and \textit{Ginex}~\cite{ginex} and \textit{Marius}~\cite{marius}~(training frameworks adapted for inference), on full-graph inference for a 2-layer SAGEConv model~\cite{hamilton2017sage} over \underline{PA}pers and \underline{MA}G-Cites (Table~\ref{tab:datasets}), on GPU workstations with $128$\,GiB RAM, RTX $4090$/$5090$ GPUs and 512\,GiB/2\,TiB SSD~(\S~\ref{sec:exp:setup}).
DGI is unable to complete the inference even within a $6$\,h time budget on MA, taking an extrapolated $\approx11$\,h for just the first layer~(Fig.~\ref{subfig:motivation-2}, left Y axis, \textit{hatched orange bars}), while the vertex-wise baseline, Ginex~\cite{ginex}, took $8$\,h~(\textit{hatched red bar}). In contrast, our \at framework completes this in $<1$\,h for both layers~(\textit{solid blue bars}).
This extends to even Marius, which uses edge-wise graph partitioning, which takes about $17$\,mins for the much smaller PA graph on the 4090 workstation~(Fig.~\ref{subfig:motivation-3}) while \at takes $\approx13$\,mins; Marius could not be evaluated on larger graphs due to lack of SSD resources on the 4090 workstation~(with 512 GiB SSD) 
and library compatibility issues with the 5090's Blackwell architecture.

\paragraph*{Proposal}

To address these challenges, we leverage a key insight: 
\textit{full‑batch GNN inference can be reformulated as a broadcast operation, enabling strictly sequential disk access instead of repeated random gathers.}
This greatly reduces read amplification. However, if done na\"{i}vely, this merely shifts the bottleneck from read I/O to aggregation and random write overheads, as intermediate outputs for all vertices may need to be materialized in memory. 
To achieve broadcast-based inference under tight memory constraints, we must \textit{avoid fully materializing intermediate embeddings while still ensuring single-pass sequential reads}.

\paragraph*{Contributions}
In this paper, we leverage this design intuition and present \at, an OOC disk-based GNN inference framework that 
enables full-graph inference on billion-scale graphs on a single machine, through broadcast‑based sequential I/O and memory-tiered aggregation.
Specifically, we make the following contributions:
\begin{enumerate}[noitemsep,topsep=2pt,leftmargin=*]
  \item \textit{Broadcast‑based inference model.} We propose a broadcast-based execution model for \at, for layer-wise GNN inference, enabling features and embeddings to be read sequentially and \textit{exactly once} per layer, significantly reducing read amplification compared to gather-based approaches.
  \item \textit{Tiered memory-disk runtime.} We design a tiered memory-disk hierarchy for \at that leverages graph topology to manage partially aggregated vertices under constrained memory, that bounds partial‑state residency, and avoids churn  
  while maintaining high throughput.
  
  \item \textit{Pipelined, overlapped execution.} We implement a full pipeline for \at that streams topology and features in chunks~(from SSD), overlaps reading with aggregation on CPU, and performs forward pass on the GPU and writing using dedicated threads, and supports multiple message-passing GNN layer architectures (GCN, GIN, GraphSAGE) with a configurable memory budget and chunk size.
  
  \item \textit{Comprehensive evaluation.} We evaluate \at on multiple large graphs ranging between 100--240 million vertices and 1.4--4\,billion edges with size on disk ranging between $54$--$550$\,GiB that easily exceed the combined RAM and GPU memory of a single workstation. 
  We demonstrate that full-graph layer-wise inference is competitive,
  taking a maximum of $\approx3$\,h for the largest graph, with substantial read I/O reduction compared to gather-based baselines. We also outperform the SOTA baselines, several of which are unable to support our largest graphs. We further present comprehensive ablation studies of all critical \at components. 
\end{enumerate}

The rest of the paper is organized as follows: we present background on GNN training and inference using gather and layer-based approaches in \S~\ref{sec:background}; we propose our \at system design in \S~\ref{sec:system}; we detail the implementation of \at, and report detailed experimental evaluation and comparative results across multiple GNNs and billion-scale graphs in \S~\ref{sec:exp}; we discuss related works in \S~\ref{sec:related}; and offer our conclusions in \S~\ref{sec:conclude}.

\section{Background}\label{sec:background}
\subsection{GNN Training and Inference}
Similar to Deep Neural Networks~(DNNs), \textit{Graph Neural Network (GNN) training} uses a two-pass~(forward and backward) approach. During the \textit{forward pass}, for a labeled vertex $u$, each layer $l$ of an $L$-layer GNN \textit{gathers} and \textit{aggregates} information from the embeddings of its neighbors. Specifically, the embeddings $\{h_v^{l-1} \mid v \in \mathcal{N}(u)\}$ from the previous layer are combined using an \textsc{Aggregate} function to produce an intermediate representation $x_u^{l}$~(Eqn.~\ref{eq:gnn-1}). This aggregated representation is then transformed by a learnable \textsc{Update} function, followed by a nonlinear activation function $\sigma(.)$, to compute the layer-$l$ embedding of vertex $u$~(Eqn.~\ref{eq:gnn-2}). This is repeated for $L$ layers, yielding the final embedding $h_u^{L}$ at the end of the forward pass. Then, a \textit{backward pass} updates the model parameters based on the loss function. 
\begin{align}
    x^l_u =&~ \textsc{Aggregate}^l(\{h^{l-1}_v, v \in \mathcal{N}(u)\}) \label{eq:gnn-1}\\
    h^l_u =&~ \sigma(\textsc{Update}^l(h^{l-1}_u, x^l_u)) \label{eq:gnn-2}
\end{align}
Here, $G=(V,E)$ is the graph, $u,v\in V$ are vertices, $l\in\{1,\dots,L\}$ is the layer index, $\mathcal{N}(u)$ denotes the in-neighbors of $u$, and $h_u^l\in\mathbb{R}^d$ is the $d$-dimensional embedding of $u$ at layer $l$.

However, recursively aggregating \textit{all} neighbors can quickly lead to out-of-memory~(OOM) errors (\textit{neighborhood explosion})~\cite{hamilton2017sage}.
To avoid this, \textit{neighborhood sampling} is performed where only a subset of neighbors of each vertex are selected to be aggregated at each hop~\cite{hamilton2017sage}.

In contrast to training, \textit{GNN inference} only requires performing a forward pass through the $L$ layers. However, while training can rely on neighborhood sampling to reduce memory costs, inference typically requires full-neighborhood aggregation to ensure accurate and deterministic vertex embeddings~\cite{inferturbo, kaler2022accelerating, ripple}. 
Consequently, full-neighborhood \textit{vertex-wise} inference~\cite{dgi, ripple} is prone to neighborhood explosion, with the frontier size increasing exponentially across layers, motivating \textit{layer-wise} inference~\cite{dgi} that computes embeddings for all vertices one layer at a time.
Layer-wise inference minimizes redundant computation, making it more scalable than vertex-wise execution. \at utilizes this layer-wise inference strategy and optimizes it for full-scale graph inference in an out-of-core setting.

\subsection{Gather-based Execution Model}
Most widely used GNNs use the \textit{message-passing} execution model, where vertex representations are iteratively updated by aggregating information from their in-neighbors. 
In \textit{gather-based} execution, each destination vertex $v$ pulls (gathers) the embeddings of its in-neighbors $\{h_u^{l-1}\mid u\in\mathcal{N}(v)\}$ to compute $h_v^l$, which induces irregular and repeated accesses when embeddings reside on disk.
Consequently, modern GNN systems primarily target message-passing GNN architectures such as GraphConv~\cite{kipf2017gcn}, GraphSAGE~\cite{hamilton2017sage}, Graph Isomorphism~\cite{gin} and Graph Attention~\cite{gat} networks.
\at optimizes the evaluation of this class of GNNs.

Broadly, message-passing GNN execution involves \textit{sample}, \textit{gather}, \textit{transfer} and \textit{compute}. During training, \textit{sample} is used to limit neighborhood explosion, where each vertex selects a random subset of its neighborhood for aggregation. 
In contrast, inference typically omits sampling to ensure deterministic \textit{full-graph} execution.
In the \textit{gather} stage, each vertex retrieves the embeddings~(or, features for the first layer) of its in-neighbors~(randomly selected or all), after which these are \textit{transferred} to the GPU. \textit{Compute} applies an aggregation function, such as \textit{sum}, \textit{mean}, \textit{max} or \textit{attention}, depending on the GNN model, to encode neighborhood information and produce updated embeddings. 

\subsection{Layer-wise Inference}%
A common approach to inference reuses training code with the backward pass disabled (\textit{vertex-wise inference})~\cite{dgi, ripple}.
However, under full-neighborhood aggregation, this approach leads to memory explosion and redundant computations, where overlapping neighborhoods are repeatedly processed.
To tackle this, layer-wise inference executes GNN computation one layer at a time by first materializing embeddings for all vertices at a given layer, and then reusing these embeddings as inputs to the next layer~\cite{dgi, ripple, inkstream}. 
This approach reduces the memory growth by processing only one layer at a time and eliminates redundant recomputation. 
However, it does not eliminate redundant data movement.
When features/embeddings are disk-resident, per-vertex gathers translate into many small, repeated reads that amplify I/O volume and reduce effective bandwidth.

This inefficiency stems from the input-unaware \textit{gather}-based execution model, where each vertex \textit{independently pulls embeddings} from its in-neighbors. Full-neighborhood aggregation induces high fan-in, causing the aggregate read volume to scale with the number of edges rather than the number of vertices.
Graph reordering techniques~\cite{chan1980linear, dgi} improve spatial locality by placing neighboring vertices closer in memory, but they do not fully eliminate redundant feature accesses.

In OOC settings, layer-wise inference still repeatedly fetches shared embeddings within a layer, causing read amplification that scales with fan-in rather than vertex count.
Since inference performs full-neighborhood aggregation with relatively lightweight computation, execution quickly becomes I/O-bound. Moreover, this access pattern repeats for each layer of the GNN, leading to extremely high cumulative read volumes.

\section{\at System Design}
\label{sec:system}

\begin{figure}[t]
    \centering
    \includegraphics[width=0.7\linewidth]{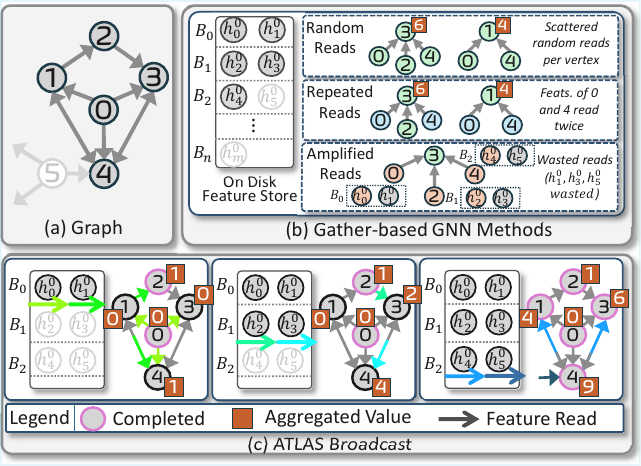}
    \caption{Illustration of gather-based versus broadcast-based execution for one layer.
    }
    \label{fig:example}
\end{figure}

\subsection{Broadcast Execution Model and Challenges}

Broadcast-based execution processes each source vertex once and pushes messages along out-edges (Fig.~\ref{fig:example}c), instead of each destination pulling from its in-neighbors (Fig.~\ref{fig:example}b).
Features are read sequentially from disk blocks~($B_0 \rightarrow B_1 \rightarrow B_2$), and each $h_u^l$ is consumed exactly once, in contrast to gather, where vertices issue scattered reads~(vertex $3$ issues reads from $\{0, 2, 4\}$), repeatedly fetch the same features~(vertex $3$ and $1$ both fetch $\{0, 4\}$), and waste already read blocks~(e.g., $h_1^0, h_3^0, h_5^0$ go unused by $3$). Broadcast replaces fan-in driven random access with fan-out driven sequential propagation~(each block $B_i$ and vertex features $h_i^0$ are read sequentially). 
Notably, this shift from gather to broadcast is semantically equivalent per layer under the same aggregation and update functions,
as seen for vertices $1$ and $3$, resulting in aggregated values of $4$ and $6$, respectively, assuming scalar vertex IDs as features and \textit{sum} as aggregator.
However, realizing broadcast-based layer-wise inference for OOC graphs poses several challenges.

\begin{enumerate}[leftmargin=0pt,itemindent=20pt,topsep=0pt,listparindent=\parindent,] 
\item \textsf{\em Bounded memory.} Full-batch, layer-wise inference places significant pressure on memory. Although broadcast-based execution avoids repeated and random reads, it produces intermediate representations for all vertices at each layer, which cannot be fully materialized in memory for large graphs.

\item \textsf{\em Execution order.} The order in which source vertices are processed determines when each destination vertex receives its neighbor's features and when it has received enough to be transformed by the GNN layer and written out. A poor order can force a large number of vertices to be partially aggregated at once, exceeding the memory budget. 

\item \textsf{\em Completion Order.} Broadcast-based execution does not enforce an order of vertex completion~(vertices that have received all the features) within a layer. Naively materializing outputs as they are produced can shift the I/O bottleneck from random reads to random writes, which can be even more expensive. 

\item \textsf{\em I/O efficiency and overlap.} Features and topology must be stored and read in a way that supports sequential access and avoids fetching the same data repeatedly. At the same time, reading data, performing aggregation, applying the layer transformation, and writing results should overlap so that the critical path is not dominated by any single stage.
\end{enumerate}

\paragraph*{Design Approach}
\at performs full-graph GNN inference using a broadcast-driven pipeline. It targets large graphs whose topology, features and intermediate embeddings do not fit in memory, executing inference directly over disk-resident data. \at is designed to avoid the limitations of traditional gather-driven layer- and vertex-wise inference under memory constraints. The system reads vertex features and topology sequentially in chunked order, propagates messages through a tiered memory hierarchy, finalizes embeddings via GPU transformation, and writes outputs as sorted spill files for the next layer. This enables \at to eliminate redundant and irregular reads caused by repeatedly fetching embeddings from storage, which otherwise leads to poor locality and high I/O amplification.
Figure~\ref{fig:atlas-arch} presents an overview of the pipeline and its components. We next describe the \at system design in detail.

\subsection{Data Layout}
We identify that data layout is a critical component of any out-of-core system. 
Accordingly, we design a simple on-disk layout for features and embeddings.
This choice is intentional, as it keeps the preprocessing step simple and time-efficient.

\begin{figure}[t]
    \centering
    \includegraphics[width=0.7\linewidth]{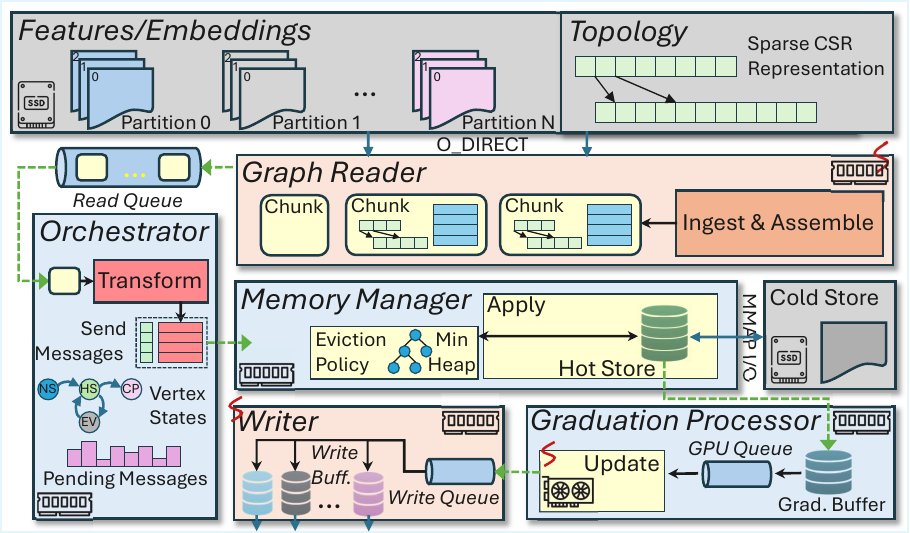}
    \caption{ATLAS architecture}
    \label{fig:atlas-arch}
\end{figure}

Since we target a broadcast-based approach for \at, messages in the form of transformed features/embeddings propagate along out-edges of vertices. For this reason, we store the graph topology in the compressed sparse row~(CSR) format~(Fig.~\ref{fig:atlas-arch}, top, \textit{green}). This representation requires $\mathcal{O}(|V| + |E|)$ space on disk and allows the \textit{graph reader} to access source vertices and edges sequentially using file offsets. 
For features and intermediate embeddings, we observe that the order in which vertices complete computation in the broadcast-based model is not sequential. As a result, storing them in a large contiguous array would require random writes, which are expensive to perform. 

Another caveat is that reordering the entire feature set to enable sequential writes would require costly external sorting due to system memory constraints.
Instead, we range-partition features and embeddings by vertex ID. 
This preserves sequential writes within each partition while avoiding a global external sort.
For each range, we maintain multiple spill files, each of which is internally sorted by vertex ID. We note that merging these spill files would again require a multi-way merge, which suffers from the same limitations discussed above. Therefore, we place the onus of presenting a sequential view of reads on the \textit{graph reader}, as discussed later.

When a group of vertices finishes computation for a layer (i.e., after all messages are received and the outputs are transformed by the GPU), their output embeddings are routed to \textit{spill buffers} associated with each range partition. Once a spill buffer is full, its contents are written to disk rather than being held until all vertices complete computation. Each spill buffer is sorted in memory by vertex ID, since it contains only a small fraction of all embeddings, and then spilled to disk. As a result, even though embeddings are produced in an arbitrary order, they can later be read in a mostly sequential manner. The graph reader can open multiple spill files for a given range and scan them sequentially to serve the next layer, without requiring a global merge or reordering step.

\subsection{Graph Reader}\label{subsec:arch-reader}
The key issue with gather-based approaches is read amplification. A vertex feature may be read multiple times, and because disk accesses occur at the block level~(typically, 4 KB), much of each block may be fetched without being used. We use this observation to design a pseudo-sequential graph reader that reads each vertex feature exactly once. This graph reader runs as a separate thread to ensure I/O can be overlapped with computation.

As noted above, features and embeddings are written as multiple spill files per range partition. The graph reader exposes a \textit{chunk}-based iterator that yields a sequence of \textit{chunks}, each corresponding to a contiguous range of vertex IDs and their corresponding features in the graph. For each chunk, the graph reader delivers two components. First, it provides the topology for the corresponding vertex range, including out-neighbors and offsets, as read from the graph's CSR representation. Second, it provides the features for the same vertices, in the same order. Each chunk is added to the \textit{reader queue} for the downstream orchestrator to consume and process. The number of vertices per chunk is configurable and determined by the user-provided size and the size of each vertex feature. 
Chunk boundaries are defined by feature bytes (not edge volume); topology is streamed from CSR for the same vertex range, so high-degree vertices increase per-chunk edge processing but do not change feature-read ordering.
Next, we describe how the graph reader creates these chunks.

For each chunk, the reader is assigned a contiguous vertex ID range [$start\_id$, $end\_id$) to read. Since features for this range may reside in multiple spill files~(as discussed previously), each spill file is indexed by its minimum and maximum vertex ID, and the list of spill descriptors is sorted by minimum ID. Using this index, the reader identifies the spill files whose ID ranges overlap the chunk's vertex range. For each relevant spill file, since the vertex IDs are stored in sorted order, this allows the reader to binary-search for the row indices corresponding to the chunk's start and end IDs and obtain a single contiguous row range. 
We then issue one aligned \textit{pread} per spill file using direct I/O~(enabled via \textit{O\_DIRECT}), bypassing the OS page cache because each feature is read exactly once and page-cache buffering provides no benefit.
Moreover, since \at targets memory-constrained systems, bypassing the page cache using \textit{O\_DIRECT} avoids cache pollution and reduces memory pressure. 
The rows retrieved from different spill files are concatenated and sorted in memory by vertex ID to produce a feature matrix in the same order as the chunk's vertex range. This \textit{merge-on-read} approach avoids a costly external merge sort over the output embeddings of all vertices at each layer. 
Spill descriptors are opened lazily, keeping open file descriptors bounded.

\subsection{Orchestrator}
The \textit{orchestrator} sits at the core of the \at system and coordinates the interaction among its components.
It maintains the system state required for execution and tracks the transformations and aggregations that must be applied to preserve GNN semantics.
It consumes chunks of topology and features/embeddings from the graph reader.
These chunks are consumed in a strictly ordered manner via a read queue populated by a dedicated reader thread, allowing disk I/O to run ahead of the main execution thread.
The orchestrator also maintains per-vertex state to track each vertex's computation progress in each layer. Upon initialization for a layer, the orchestrator coordinates and initializes other components, including the memory manager and the graduation processor.
It serves as the entry point for executing the current layer and requests both components to perform their required memory allocations. 
Additionally, it initializes the data structures associated with the selected eviction policy.
It creates a \textit{pending messages} tracker for every vertex to record how many messages the vertex has received cumulatively before the processing of the current chunk.
Next, it also records a vertex's state based on the number of messages it has received. Each vertex can only be in \textit{exactly one} of the four possible states. 

A vertex is in the \textit{NOT STARTED} state if it has not yet received any messages for the current layer. It transitions to the \textit{HOT} state once it starts receiving messages and its partial aggregation state is resident in memory, as we discuss next. If the buffer holding this partial state is evicted due to memory pressure before the vertex has received all required messages, the vertex enters the \textit{COLD} state, and its intermediate state is spilled to disk.
Finally, a vertex reaches the \textit{COMPLETED} state once it has received all messages for the layer. We note that the only valid state transitions are from \textit{NOT STARTED} to \textit{HOT}, from \textit{HOT} to either \textit{COLD} or \textit{COMPLETED}, and from \textit{COLD} back to \textit{HOT} when the spilled state is reloaded for continued processing. No other transitions are permitted.
The global orchestrator state is stored as compact per-vertex arrays (e.g., degrees, pending counts, and 1-byte state), requiring $\mathcal{O}(|V|)$ memory and remaining within a few GiB even for IF-scale graphs~(Tab.~\ref{tab:datasets}).

Next, to connect the orchestrator's and other \at components' execution with GNN semantics, we describe the mathematical formulation of a layer with \textit{mean} aggregation. Let $G = (V, E)$ be a directed graph. Let $h_u^l \in \mathbb{R}^d$ denote the embedding of vertex $u$ at layer $l$, and let $\mathcal{N}(v)$ denote the set of in-neighbors of vertex $v$. The output embedding of vertex $v$ at the next layer is given by
$h_v^{(l+1)} = \sigma \left( W^{(l)} \cdot \frac{1}{|\mathcal{N}(v)|} \sum_{u \in \mathcal{N}(v)} h_u^{(l)} \right)$, 
where $W^{(l)}$ is the learnable weight matrix for layer $l$, and $\sigma(\cdot)$ is a non-linear activation function. During execution, the orchestrator implements this formulation in a broadcast-based manner. For every outgoing edge $(u, v)$, the orchestrator constructs a message, $m_{u \rightarrow v}^{(l)} = \frac{1}{|\mathcal{N}(v)|} \, h_u^{(l)}$,
where normalization by the destination degree is applied at message construction time. These messages are emitted as records of the form $\langle v, m_{u \rightarrow v}^{(l)}, s_v \rangle$, where $s_v$ denotes the current state of the destination vertex. These records are forwarded to the memory manager, which uses the vertex state to determine whether to assign, update, or reload the aggregation buffer for a vertex into the \textit{hot store}.

\subsection{Memory Manager}
The memory manager holds vertex partial aggregated state for the current layer under a fixed configurable RAM budget. It manages a RAM--disk hierarchical store to hold the partial states of vertices that are currently active, i.e., vertices that have received $\geq 1$ messages but not all of them. It also manages a configurable \textit{eviction policy} to decide which vertices get a slot in the RAM~(are maintained in the \textit{hot store}) and which vertices need to be evicted to the \textit{cold store} on disk. Specifically, the memory manager is responsible for a \textit{hot store} in RAM, a \textit{cold store} on disk and the eviction policy to decide which vertices need to transition from \textit{HOT} to \textit{COLD}. 

\subsubsection{Hot Store} The \textit{hot store} is a fixed-size array of slots, allocated upon initialization, with each slot holding the partial state of a single vertex. It also aggregates incoming messages from the orchestrator directly into these slots using a vertex-to-slot mapping. 
Slots are assigned on demand when vertices transition to \textit{HOT} and are freed once vertices receive all messages and move to \textit{COMPLETE}.
When the hot store has no free slots and additional vertices must be brought in, the memory manager consults the eviction policy and evicts some of the existing vertices to the \textit{cold store}. The cold store is a disk-backed tier on SSD, and it preserves the feature state of evicted vertices so that they can be reactivated later when they are again required as message destinations. 
We note that a vertex can update its partial state only while it resides in the hot store.
As such, the \textit{cold store} serves solely as backing storage for vertices that have been evicted and are waiting to be reactivated into the hot store. The memory manager also interacts with the orchestrator, keeping it informed about the state changes of any vertex. 

\subsubsection{\at Eviction Policy} When the hot store exhausts its available slots, and evictions are required, the choice of vertices to evict is delegated to an eviction policy. System performance can be bottlenecked by frequent disk reads incurred when partial states are repeatedly reloaded from the cold store into the hot store, or by repeated evictions to the cold store. If vertices are evicted at random, a vertex may undergo multiple eviction-and-reload cycles before completing aggregation, significantly degrading performance. 
To circumvent this, \at employs a \textit{minimum-pending-messages} eviction policy. 
Vertices with the fewest remaining pending messages are selected for eviction, as they are closer to completion and are more likely to complete the next time they are reloaded to the hot store. This reduces repeated \textit{eviction-reload cycles} of the partial states of the same vertex and limits unnecessary disk traffic. 

The eviction policy must maintain an ordered view of vertices currently in the hot store. Additionally, it should support insertion, removal, score updates as messages arrive, and selection of the k-smallest vertices for eviction. We implement this using a custom bucket-based min-heap, rather than Python's \texttt{heapq}. We exploit the fact that eviction scores are integers in a small bounded range~($[0, max\_in\_degree]$), corresponding to pending message counts. Vertices are stored in score-indexed~(pending messages-indexed) buckets implemented as doubly linked lists, allowing constant-time insertion, removal, and score decrement. Eviction selects victims by scanning the smallest non-empty buckets, resulting in $\mathcal{O}(k)$ for choosing $k$ vertices.

\subsubsection{Cold Store} The cold store is implemented as a memory-mapped file using Numpy's \texttt{mmap} API. This design choice deviates from our earlier decision to avoid the page cache using $O\_DIRECT$. However, this is done intentionally. An evicted vertex is guaranteed to be reloaded into the hot store in a subsequent chunk, making page-cache buffering beneficial rather than wasteful. By using buffered I/O, we exploit caching effects to reduce the cost of repeated reads and writes for evicted vertex states.

Lastly, the memory manager is also responsible for offloading vertices that have received all messages to the \textit{graduation processor}. We discuss this next.

\subsection{Graduation Processor}
The \textit{graduation processor} is responsible for processing vertices that have received messages from all of their in-neighbors. 
When a vertex's pending message count reaches zero, the orchestrator signals the memory manager to finalize aggregation and release the corresponding hot store slots. The finalized aggregated features are copied into a configurable \textit{graduation buffer} managed by the graduation processor. This allows hot store slots to be freed immediately. 

Once the graduation buffer is full, it is offloaded to the \textit{GPU queue}~(Fig.~\ref{fig:atlas-arch}). To avoid stalling chunk execution in the main thread, the graduation processor uses \textit{double buffering}, where one buffer is offloaded while the main thread continues writing graduated vertices into the alternate buffer.
The graduation processor also launches a dedicated GPU offload thread that dequeues buffers from the GPU queue and transfers the aggregated vertex states to the GPU.
The linear transformation and non-linearity are applied on the GPU, and the thread waits for the results to be transferred back to host memory. By offloading this work to a separate thread, the graduation processor avoids blocking on GPU execution or data transfers. 
The GPU offload thread uses CUDA streams to overlap data movement and computation, and once the transformed embeddings are available, it enqueues them to the write queue for disk output.

\subsection{Embedding Writer\label{subsec:method-emb-writer}}
Finally, the transformed embeddings produced by the graduation processor must be written to disk so they can be consumed by the next layer or used as final output. This is handled by the writer that runs in a dedicated writer thread and consumes batches of (vertex IDs, transformed embeddings) from the \textit{write queue}. Vertices arrive at the writer in graduation order, which is arbitrary with respect to the order of processing and is, therefore, unsuitable for downstream consumption as is. To address this, the writer range-partitions vertices by ID, similar to how the reader expects the inputs, ensuring that each partition corresponds to a disjoint vertex ID range.

For each partition, the writer maintains in-memory \textit{spill buffers} (Fig.~\ref{fig:atlas-arch}) for vertex IDs and embeddings. Incoming batches are scattered into these per-partition spill buffers. When a buffer fills, its contents are sorted by vertex ID and flushed to disk as a sorted spill file. Over the course of a layer, the writer may produce multiple sorted spills per partition. As also previously discussed in \S~\ref{subsec:arch-reader}, this design avoids performing a global merge or maintaining a fully sorted output during layer execution, which would be expensive under memory constraints. The writer, like the reader, also employs a non-buffered I/O strategy and uses $O\_DIRECT$ to write aligned buffers to the spill files. This helps \at perform large, contiguous writes to disk, avoiding system-call overhead while also preventing page-cache pollution from finalized embeddings that are not going to be reused in the current layer.

\subsection{Graph Reordering}\label{subsec:order}
Proceeding with layer-wise execution in the original vertex ID order might not be optimal and can lead to memory pressure, increased evictions, reduced graduation throughput, and overall system slowdown. Vertices might occupy the hot store during initial phases and might not complete till the last chunk is processed, resulting in increased eviction frequency. 

Before executing layer-wise inference, \at provides the option to reorder the graph by reassigning vertex IDs and relabeling the topology and features accordingly. For downstream layers, \at graph reader follows this new order while producing the chunks for processing. We note that while reordering is a one-time offline cost that amortizes across layer executions, it is not trivial to do so in a memory-constrained setting. Next, we describe the approach to this reordering scheme. \at reorders the graph while keeping two constraints in mind; to \textit{maximize the vertex completion rate} while \textit{minimizing the number of vertices that remain partially aggregated} at any point in time. To achieve this, \at adopts a greedy message-passing heuristic to compute the vertex ordering. Next, we define this formally.

Let $G = (V, E)$ be a directed graph. For a vertex $v \in V$, let $d_{\text{in}}(v)$ and $d_{\text{out}}(v)$ denote its in-degree and out-degree, respectively. Let $k_v(t)$ denote the number of messages received by vertex $v$ after processing the first $t$ vertices in a given ordering. Therefore, a vertex $v$ is complete when 
$k_v(t) = d_{\text{in}}(v)$.

We define the fractional completion state of a vertex at $t$ as
$c_v(t) = \frac{k_v(t)}{d_{\text{in}}(v)}$.
The global completion state at $t$ is
$\phi(t) = \sum_{v \in V} c_v(t) = \sum_{v \in V} \frac{k_v(t)}{d_{\text{in}}(v)}$.
When the next vertex $u$ is processed at $t+1$, it sends one message to each vertex $v \in \text{Out}(u)$, increasing their message counts by one:
$k_v(t+1) = k_v(t) + 1, \quad \forall v \in \text{Out}(u)$.
Therefore, the marginal gain in the completion state can be given by $\Delta \phi(u) = \phi(t+1) - \phi(t)
= \sum_{v \in \text{Out}(u)} \frac{1}{d_{\text{in}}(v)}$.
So, we should pick a vertex $u$ that provides the maximal marginal gain to process next. 

However, to balance completion benefit against memory requirements for processing, \at assigns each vertex $u$ a score of
$\text{Score}(u) =
\frac{\sum_{v \in \text{Out}(u)} \frac{1}{d_{\text{in}}(v)}}{d_{\text{out}}(u)}$,
where the denominator is a heuristic of how many buffers a vertex might open if it is picked. This is the cost of picking vertex $u$.
Vertices are ordered greedily in decreasing order of this score, favoring vertices whose processing contributes most to global completion while emitting fewer messages. 
This strategy has the benefit of being light-weight enough to be calculated in only a single pass over the graph topology information. 

Finally, \at reads the original feature matrix in old-ID order and processes it in chunks. Each vertex ID in a chunk is first mapped to its new ID using the relabeling map, and then assigned to an output partition based on the new ID, using the same range-partitioning scheme employed by the runtime writer. Features are buffered per partition, sorted by new ID within each buffer, and written to disk as partitioned spill files. As a result, feature vectors are laid out on disk in increasing new-ID order within each partition, matching the access pattern expected by the graph reader during layer execution.

\subsection{Generalizability of \at}
Lastly, while \at currently only supports specific GNN architectures, we note that any other message-passing GNN architecture, such as GAT~\cite{gat} or SGC~\cite{sgc}, can be easily adopted into this execution model. Alternative GNN semantics, such as neighborhood sampling, can also be supported by running a preprocessing pass over the topology to disable edges that are not required, thereby simulating sampling. However, since most GNN architectures combine a vertex's own embeddings with the aggregated embeddings of its in-neighbors, it is important to note that all vertex features still need to be read from disk, making \at and its optimizations highly relevant.

\section{Experiments and Results}
\label{sec:exp}

\begin{table}[t]
    \centering
    \caption{Graph datasets~\cite{ogb, igb} used in experiments.}
    \label{tab:datasets}
    \footnotesize
    \begin{tabular}{c|c|c|c|c}
        \hline
        \textbf{} & \textbf{Papers} & \textbf{MAG-Cites} & \textbf{IGB-Large} & \textbf{IGB-Full} \\ \hline\hline
        \textbf{Abbr.} & PA & MA & IL & IF \\ \hline
        \textbf{Vertex Count} & 111M & 121M & 100M & 269M \\ \hline
        \textbf{Edge Count} & 1.7B & 1.4B & 1.2B & 4B  \\ \hline
        \textbf{Feat. Dim} & 128 & 768 & 1024 & 1024 \\ \hline
        \textbf{Num. Classes} & 172 & 153 & 19 & 19 \\ \hline
        \textbf{Topology Size~(GiB)} & 27 & 22 & 19 & 56 \\ \hline
        \textbf{Feature size~(GiB)} & 54~(FP32) & 175~(FP16) & 200~(FP16) & 550~(FP16) \\ \hline
    \end{tabular}
\end{table}

\subsection{Experimental Setup}\label{sec:exp:setup}
We use three popular GNN models for the vertex classification task -- GraphConv~\cite{kipf2017gcn}, SAGEConv~\cite{hamilton2017sage}, and GINConv~\cite{gin} -- with 2 layers and hidden dimension set to $128$. To evaluate \at, we use four popular open-source large-scale datasets described in Tab.~\ref{tab:datasets}. These feature sets of these datasets range between medium-sized~($54$\,GiB for \textit{Papers}~\cite{ogb}) with FP32 precision to large-scale for \textit{IGB-full}~\cite{igb} at $550$\,GiB using FP16 representation. We perform full-graph GNN inference where no incoming edges for any vertex are sampled, i.e., all neighbors participate at each hop of the computation graph. 
We note that \at produces output representations that match the reference within the floating-point precision bounds. 
The reference is an in-memory full-batch layer-wise implementation with identical weights and precision on PA using DGL~\cite{wang2019deep}.
For PA, the \emph{mean over vertices of the maximum absolute error across output dimensions} is $8\times10^{-5}$, while the \emph{mean over vertices of the average relative error across output dimensions} is $2.8\times10^{-6}$.

All our experiments are performed on a single GPU workstation with a $12$-core AMD Ryzen $9$ $9900$X processor~($4.4$ GHz), $128$\,GiB of RAM, an NVIDIA RTX $5090$ GPU card with $32$\,GiB of memory. The workstation is also equipped with a $2$ TiB Samsung $990$ PRO SSD and runs Ubuntu $24.04.3$ LTS. 
To eliminate OS page-cache effects across runs (especially for mmap-based baselines), we clear the page cache at the start of each experiment.
Lastly, we obtain per-process I/O statistics, including bytes read and written, from \texttt{/proc/<pid>/io}.

\subsection{\at Implementation and Baselines}
\at is implemented in Python using NumPy v2.0 and PyTorch v2.8. The \textit{graph reader} and \textit{embedding writer} are implemented in C++ and integrated via PyTorch C++ extensions, compiled using \texttt{ninja}. 
The framework overhead can vary depending on the sizes of the graduation buffers at the graduation processor, the spill buffers at the \textit{writer}, the maximum queue sizes, etc. For our experiments, with a chunk size of 8 MiB, graduation buffer of size 256 MiB, 8 spill buffers of size 1 GiB each~(total 8 GiB), all queues of size 20, with the framework overhead across datasets varying between 6--7 GiB.

We implement \at~(\textbf{AT}) against two key baselines for comparison, one based on vertex-wise execution~(Ginex~\cite{ginex}, \textbf{GN}) and the other based on layer-wise execution~(DGI~\cite{dgi}, \textbf{DG}). Ginex is primarily designed for GNN training. It builds and stores a neighbor cache on disk for a vertex-based popularity score and performs \textit{sampling} of computation graphs across multiple batches of training vertices~(called \textit{superbatch}), using this neighbor cache. 
We modify Ginex to perform only the forward pass to simulate inference and use the default superbatch and batch sizes of 2500--3300 and 1000, respectively, from the paper. Since our testbed machine has 128 GiB of RAM, we allocate 80 GiB for the feature cache and 10 GiB for the neighbor cache.

DGI, on the other hand, is a layer-wise full-graph inference baseline that proposes dynamic batching and graph reordering to optimize inference. DGI maps the features and CSC indices on disk as NumPy \texttt{mmap} files and uses buffered I/O for disk reads and writes. Moreover, they use the RCMK ordering~\cite{chan1980linear} for optimal results. We adopt the settings from the paper as is for DGI. 

We report layer-wise results for both DGI and \at and consolidated results for Ginex since it follows a vertex-based execution model. Lastly, we run each experiment for up to $6$\,h and record the percentage of execution completed within that time. The final time is then extrapolated from this linearly using the fraction of the layer completed within $6$\,h (by processed vertex range/chunks).
This is done at a layer level for DGI and \at, and at the framework level for Ginex. For our experiments, we assign the hot store as $50$\,GiB, $70$\,GiB, $80$\,GiB, and $100$\,GiB for PA, MA, IL, and IF, respectively.
We choose the smallest hot-store size that~(nearly) eliminates evictions for each dataset, capped at 100\,GiB to leave headroom for other framework buffers and overhead.

\begin{figure}[t]
    \centering
    \subfloat[PA/GCN2]{
        \includegraphics[width=0.24\columnwidth]{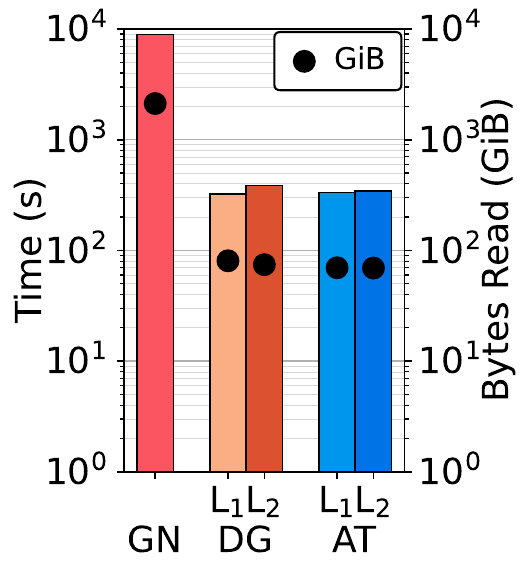}
    }
    \subfloat[PA/SAGE2]{
        \includegraphics[width=0.24\columnwidth]{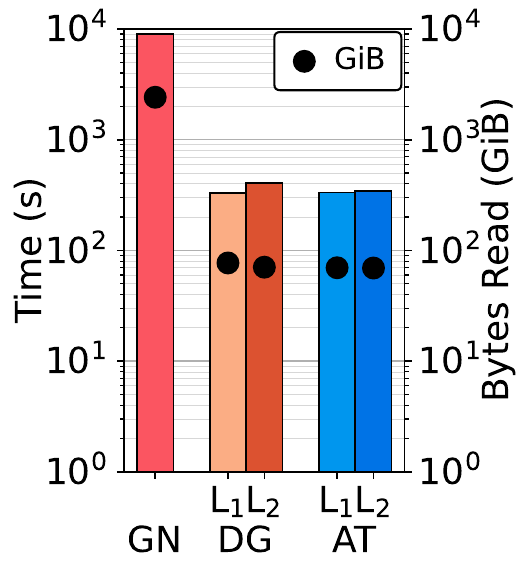}
    }
    \subfloat[PA/GIN2]{
        \includegraphics[width=0.24\columnwidth]{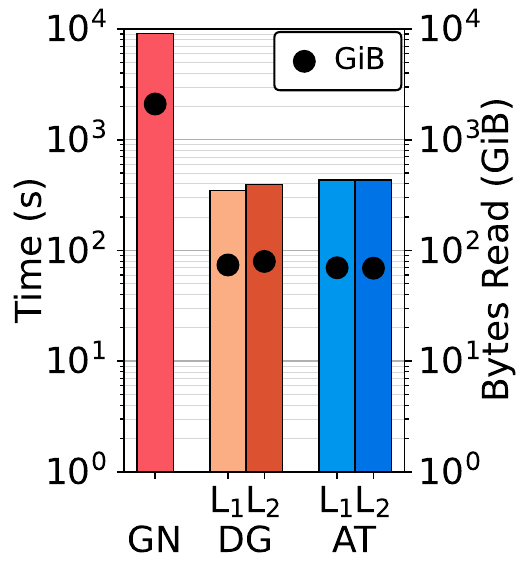}
    }
    \subfloat[MA/GCN2]{
        \includegraphics[width=0.24\columnwidth]{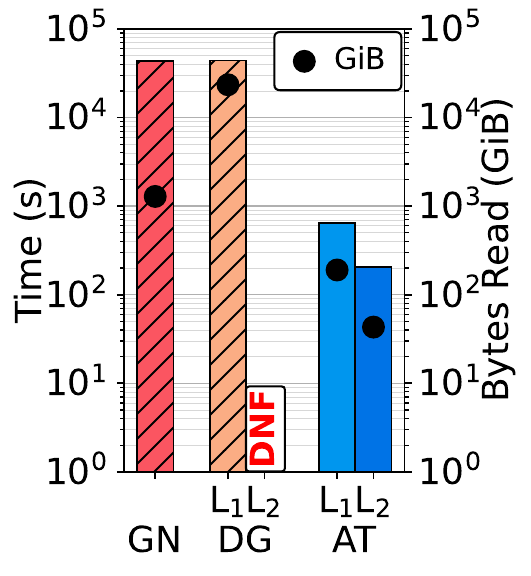}
    }\\
    \subfloat[MA/SAGE2\label{ma-sage}]{
        \includegraphics[width=0.24\columnwidth]{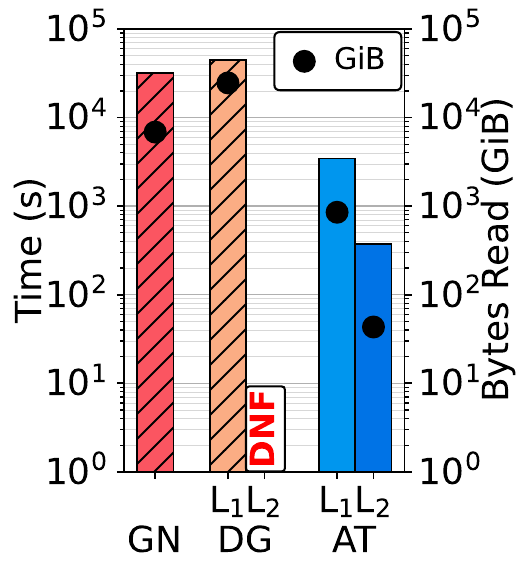}
    }
    \subfloat[MA/GIN2]{
        \includegraphics[width=0.24\columnwidth]{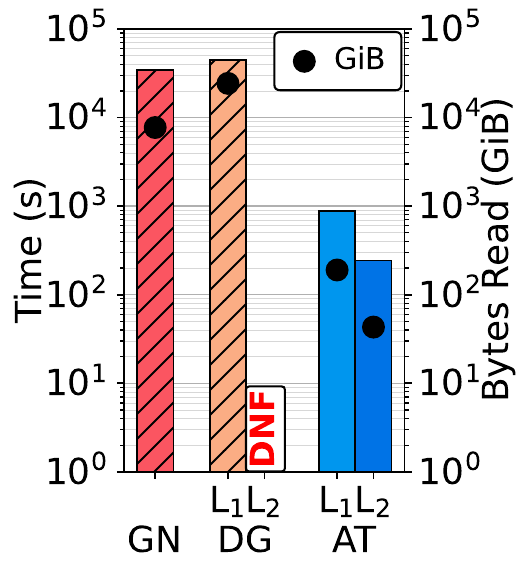}
    }
    \subfloat[IL/GCN2]{
        \includegraphics[width=0.24\columnwidth]{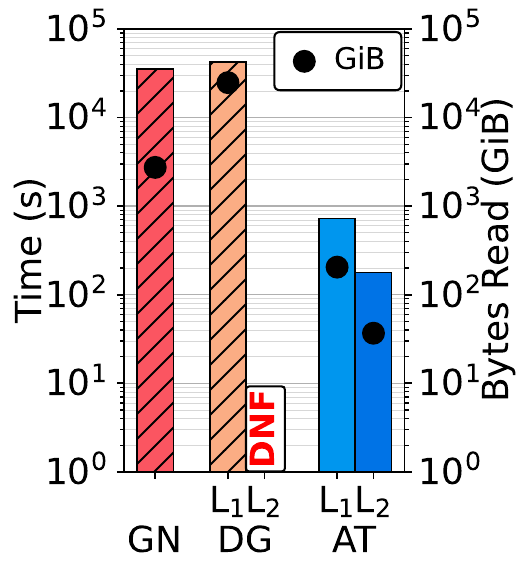}
    }
    \subfloat[IL/SAGE2\label{il-sage}]{
        \includegraphics[width=0.24\columnwidth]{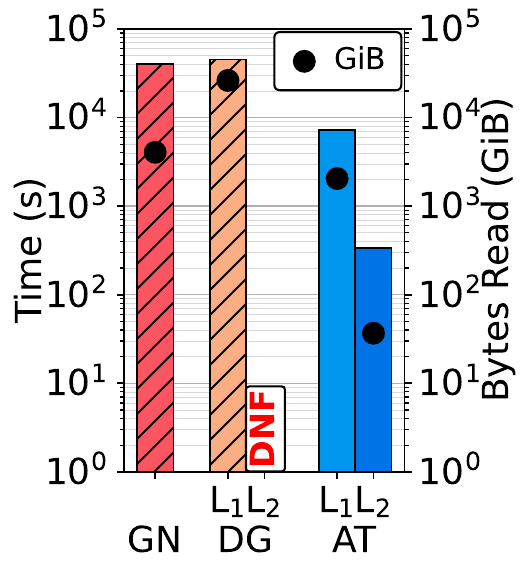}
    }\\
    \subfloat[IL/GIN2]{
        \includegraphics[width=0.24\columnwidth]{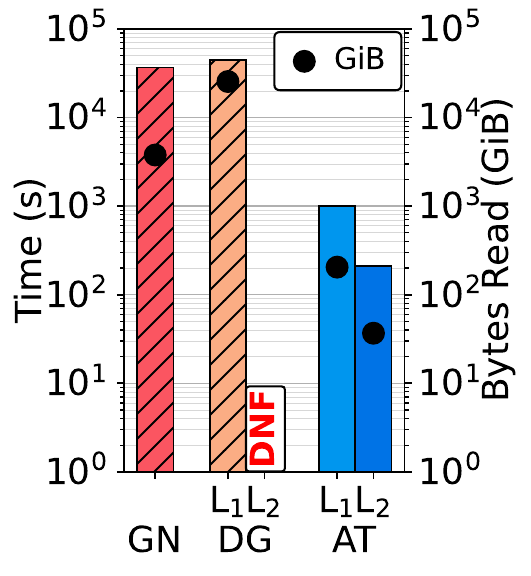}
    }
    \subfloat[IF/GCN2]{
        \includegraphics[width=0.205\columnwidth]{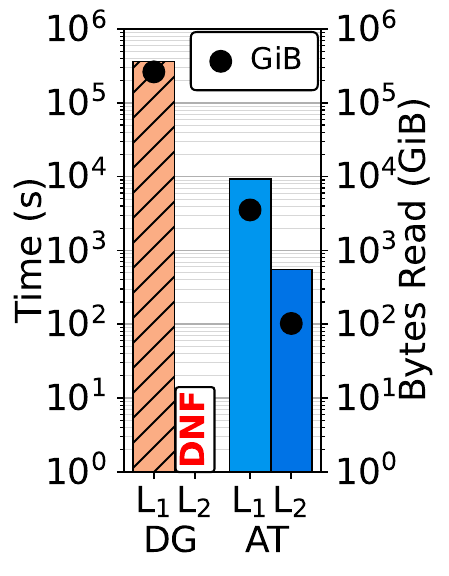}
    }
    \caption{Time taken~(left Y axis, \textit{hatched bars} for extrapolated, \textit{solid bars} for completed runs) and extrapolated data read from disk~(right Y axis, \textit{marker}) for complete execution of Ginex and layer-wise executions of DGI and \at.}
    \label{fig:baseline-comparison}
\end{figure}

\subsection{Comparison with SOTA Baselines}
Fig.~\ref{fig:baseline-comparison} reports the total/extrapolated time of execution~(left Y axis) of 2-layer GNN models across all 4 datasets for the 3 frameworks. 
For Ginex, each bar is end-to-end time; for DGI and \at, bars report per-layer time and the total inference time is the sum across layers (when both complete).
The executions that did not run are marked as ``DNF''.
Across OOC datasets (MA/IL/IF), \at reduces disk traffic by 1--2 orders of magnitude relative to gather-based baselines, yielding large runtime improvements; on PA (fits in-memory), 
\at matches DGI within $\approx5\%$ runtime overall, while reducing disk traffic by $\approx9\%$.

\paragraph{Performance Improvements over Ginex} We observe an average decrease in execution time for \at over Ginex of $12.4\times$, $30\times$, and $23.4\times$ for PA, MA and IL, with the latter two being extrapolated times. This is primarily because Ginex does multiple I/O operations before and within each superbatch. Precisely, during each superbatch sampling phase, Ginex loads the neighbor cache from SSD and writes all batches within the superbatch to disk. Within the processing phase, the feature cache is initialized from disk, and even during batch processing, Belady's algorithm reduces the I/O but does not eliminate it. In contrast, \at does sequential reads and writes \textit{exactly once}. Additionally, the vertex-wise inference baseline also tends to perform redundant computation during the forward pass. However, we see from the extrapolated amount of data read by Ginex that I/O is the major bottleneck. Ginex reads on average $16\times$, $15\times$, $11\times$ more data than \at~(combined for both layers) for PA, MA, and IL, respectively, with absolute values going as high as $7.7$\,TiB on MA~(feature size 175\,GiB).

\paragraph{Performance Improvements over DGI} For DGI, we note that, since the first-layer executions in Fig.~\ref{fig:baseline-comparison}(d)--(j) did not finish within the allocated 6 hours, the second-layer execution did not start. For PA, from Fig.~\ref{fig:baseline-comparison}(a)--(c), we see that both DGI and \at exhibit similar performances within $5\%$ of each other. This is because the PA graph fits entirely in the memory of our testbed machine with 128 GiB of RAM. When DGI begins execution, 
the buffered behavior of \texttt{mmap}
causes the accessed blocks to be paged into memory. Subsequent accesses are then served from the page cache, effectively avoiding disk I/O. However, for MA and IL, the feature sizes are too large to fit in memory. This can also be observed from the extrapolated time taken by $L_1$ of DGI being $44\times$ and $36\times$ more than AT's $L_1$ time for MA and IL, respectively, averaged across GNN models.

\paragraph{Increase in \at execution time for SAGEConv for MA and IL}
We note a sharp increase in the $L_1$ execution times for SAGEConv as compared to GCN2 and GIN2 for both MA~(Fig.~\ref{ma-sage}) and IL~(Fig.~\ref{il-sage}), increasing from an average of $\approx800$s to $3400$s for MA and $\approx7000$s for IL. This is because the SAGEConv model concatenates its own embeddings with the aggregated embeddings of its in-neighbors. This doubles the number of columns in the hot store per vertex, effectively halving its size. This, in turn, leads to a rise in evictions and partial states being read or written to or from the disk. This can also be observed in the higher number of bytes read in Fig.~\ref{ma-sage} and Fig.~\ref{il-sage} compared to the other 2 models. However, even with this restriction, AT's $L_1$ outperforms DGI's $L_1$ by $13\times$ for MA and $6.3\times$ for IL.

\paragraph{Impact of hidden dimension on layer execution time} We notice the layer execution time for \at dropping~from $L_1$ to $L_2$ for both MA and IL by $\approx5\times$ and $10\times$ on average, respectively. This is primarily because both MA and IL have high feature dimensions of $768$ and $1024$, respectively, whereas hidden dimension sizes in most GNN architectures range from $32$ to $256$. In our case, the hidden dimension is set to $128$. 

\paragraph{Performance Improvement over DGI on IF}
Finally, we show the performance of \at on the largest dataset IF. Note that Ginex could not be evaluated on this dataset because of runtime errors. We see a massive reduction in execution time of $\approx39\times$ for $L_1$ of \at vs $L_1$ of DGI. This is again due to the memory pressure created by large, repeated, random reads in DGI, as evidenced by the extrapolated read size of $\approx262$ TiB for $L_1$, compared to \at's $3.6$ TiB summed across both layers.

\subsection{Resource Utilization of \at}
\begin{figure}[t]
    \centering
    \subfloat[IL/GCN2 CPU and Memory\label{subfig:il-cpu-mem}]{
        \includegraphics[width=0.35\columnwidth]{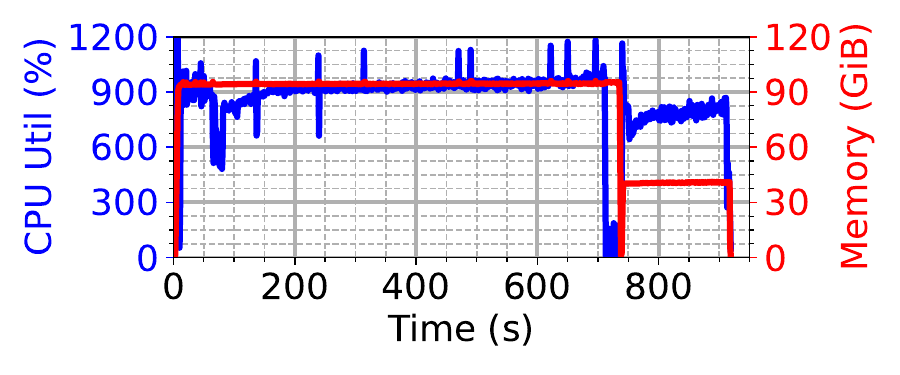}
    }%
    \subfloat[MA/GCN2 CPU and Memory\label{subfig:ma-cpu-mem}]{
        \includegraphics[width=0.35\columnwidth]{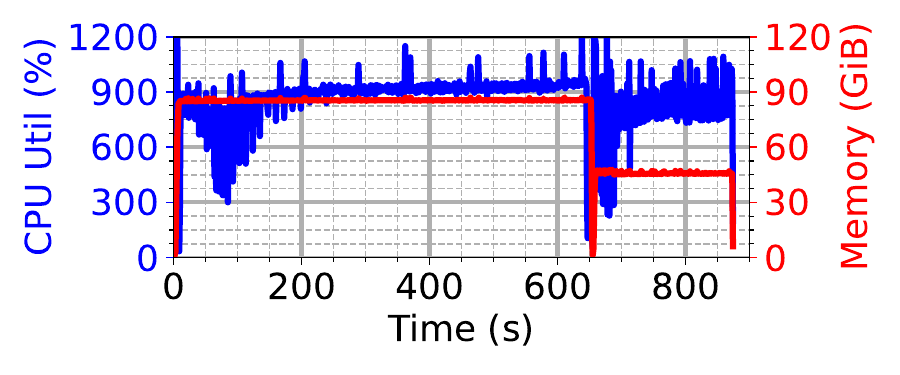}
    }\\
    \subfloat[IL/GCN2 GPU Util and Memory\label{subfig:il-gpu-mem}]{
        \includegraphics[width=0.35\columnwidth]{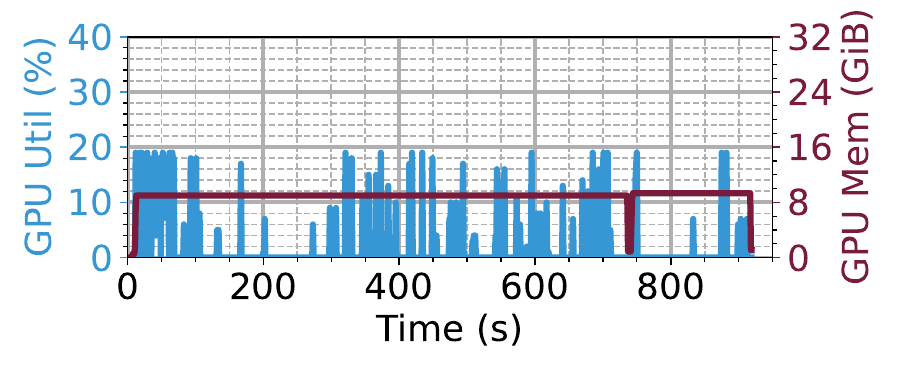}
    }%
    \subfloat[MA/GCN2 GPU Util and Memory\label{subfig:ma-gpu-mem}]{
        \includegraphics[width=0.35\columnwidth]{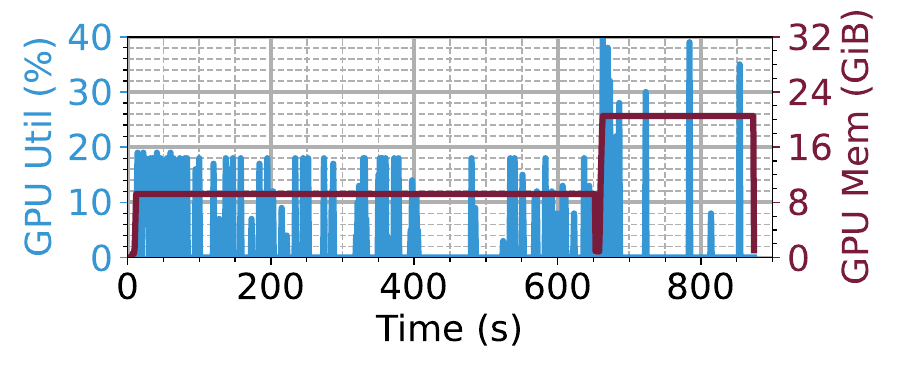}
    }\\
    \subfloat[IL/GCN2 Read/Write Bandwidth\label{subfig:il-rw}]{
        \includegraphics[width=0.35\columnwidth]{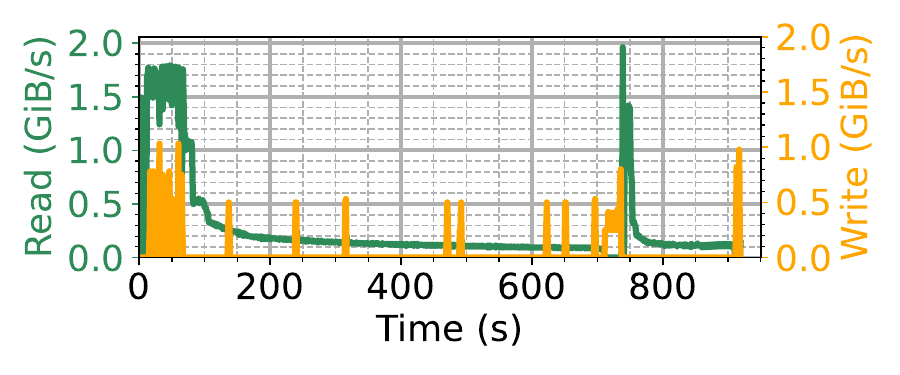}
    }%
    \subfloat[MA/GCN2 Read/Write Bandwidth\label{subfig:ma-rw}]{
        \includegraphics[width=0.35\columnwidth]{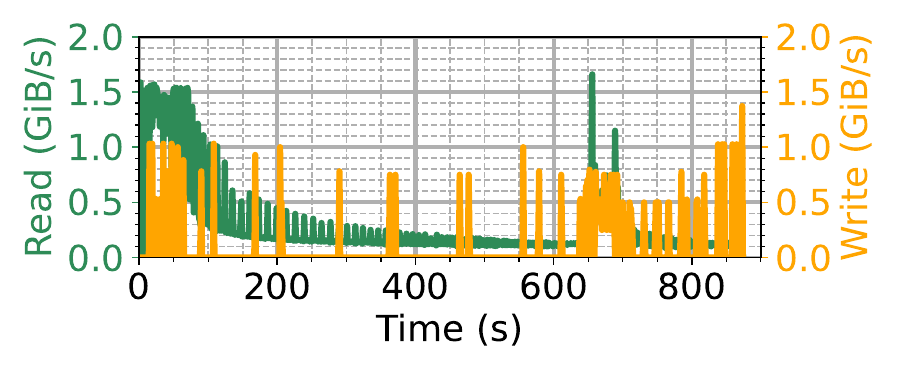}
    }\\
    \subfloat[IL/GCN2 Read-Only microbenchmark vs. \at\label{subfig:read-only-vs-atlas}]{
        \includegraphics[width=0.18\columnwidth]{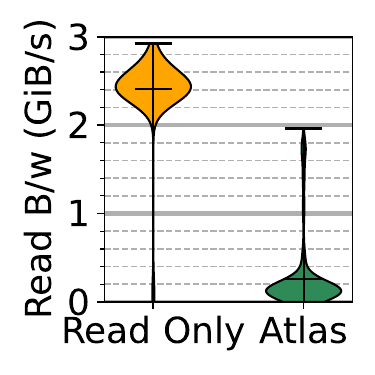}
    }
    \caption{Resource utilization for a 2-layer GCN on the IL and MA datasets. \textit{First row: } CPU~(\textit{blue}, left Y axis) and memory~(\textit{red}, right Y axis) over time. \textit{Second Row: } GPU util~(\textit{sky blue}, left Y axis) and memory~(\textit{red}, right Y axis) usage over time.
    \textit{Third row:} SSD read~(\textit{green}, left Y axis) and write~(\textit{orange}, right Y axis) bandwidth over time. \textit{Last row: }Baseline ``read-only'' microbenchmark vs. ATLAS active read rate.}
    \label{fig:util}
\end{figure}
Fig.~\ref{fig:util} reports CPU utilization, resident memory, and SSD read/write bandwidth, samples at a 1 second interval, over a complete 2-layer GCN run on IL and MA datasets with hot-memory budgets set to 70 GiB for MA and 80 GiB for IL.

\paragraph{Compute} Figs.~\ref{subfig:il-cpu-mem} and~\ref{subfig:ma-cpu-mem} show that ATLAS sustains a high and stable CPU utilization~(\textit{blue line}, left Y axis) throughout both dataset runs, with an average of $856\%$ for IL and $852\%$ for MA on a 12-core machine~(\S~\ref{sec:exp:setup}). The sustained high CPU utilization indicates that the computation threads are not bottlenecked by the upstream reader or downstream writer I/O tasks, allowing the system to efficiently execute the dense feature aggregation operations. We notice a sharp dip in the CPU utilization for both IL and MA around the $\approx700$ second and $\approx650$ second marks, respectively. This is the layer boundary as the \at orchestrator prepares to process the next layer.

\paragraph{Memory}
The memory footprint~(\textit{red lines}, right Y axis) remains stable at $\approx95$\,GiB during Layer~1~($0$--$700$ seconds) for IL. This includes $80$\,GiB for the hot store, $8$\,GiB for spill buffers~(\S~\ref{subsec:method-emb-writer}), and an additional $\approx6$--$7$\,GiB for \at framework overheads. We see a similar trend for MA during Layer~1~($0$--$650$ seconds), where memory usage stabilizes at $\approx85$\,GiB, consisting of $70$\,GiB for the hot store, $8$\,GiB for spill buffers, and the remainder for framework overheads. The memory footprints drop significantly for Layer~2 across both datasets to $40$\,GiB for IL and $\approx47$\,GiB for MA as the intermediate embeddings produced by Layer~1~(128 dim) are much smaller than the raw input features~(e.g., 1024 for IL), and the total hot-store budget is not utilized. Notably, the memory overhead from spill buffers and other framework components still persists.

\paragraph{GPU Utilization}
Figs.~\ref{subfig:il-gpu-mem} and~\ref{subfig:ma-gpu-mem} demonstrate the GPU utilization~(\textit{sky blue}, left Y axis) and memory usage~(\textit{brown}, right Y axis) during complete \at runs. We notice \at exhibits intermittent bursts of GPU compute, peaking at $\approx20\%$ during Layer~1 for both datasets, and reaching up to $40\%$ during Layer~2 for MA.
The bursty utilization reflects the offload of full \textit{graduation buffers} by the graduation processor. The GPU memory used, in addition to fixed overheads of model weights and biases and the Pytorch and CUDA context, are proportional to the offloaded buffer size as well as the output dimensions for that layer. The used GPU memory for IL holds steady at $\approx9$\,GiB across both layers. However, we observe that GPU memory increases from $\approx9$\,GiB to $\approx20$\,GiB for Layer~2 of MA, accommodating the larger output dimension of $153$.

\paragraph{I/O Bandwidth} 
Figs.~\ref{subfig:il-rw} and~\ref{subfig:ma-rw} show the SSD read (\textit{green lines}, left Y axis) and write~(\textit{orange lines}, right Y axis) bandwidth during execution. 
We observe sharp read spikes~(up to $\approx1.5$--$1.8$\,GiB/s) at the start of execution and at layer boundaries for both datasets, corresponding to the initial filling of the read queues. 
Additionally, the \at ordering front-loads vertices with very small in-degree (e.g., only self-loops), allowing them to complete quickly and, therefore, exhibits high upstream read throughput early in the layer execution. 
However, during the bulk of the layer execution, the sustained read bandwidth remains relatively low, averaging $\approx0.25$\,GiB/s for both IL and MA, indicating the execution is relatively \textit{compute-bound} and that read rate throttle is due to read queue backpressure. 
We also validate this through Fig.~\ref{subfig:read-only-vs-atlas}, which compares \at's active read rate against a sequential ``read-only'' microbenchmark on the same SSD. While the NVMe SSD hardware can sustain $\approx2.5$\,GiB/s for reading chunks without any downstream processing, \at utilizes only $\approx0.25$\,GiB/s on average during execution. 
Finally, the intermittent write spikes correspond to the periodic flushing of the \textit{full} in-memory spill buffers to disk as vertices graduate. We again observe dense groupings at the start of both layers, suggesting the rapid graduation of very low in-degree vertices at the start of execution.

\subsection{Impact of \at Ordering}
\begin{figure}[t]
    \centering
    \subfloat[IL/GCN2\label{subfig:reorder-1}]{
        \includegraphics[width=0.4\columnwidth]{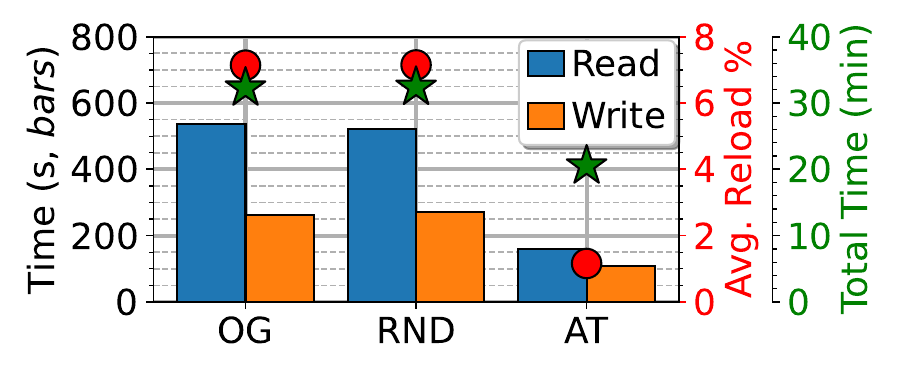}
    }%
    \subfloat[MA/GCN2\label{subfig:reorder-2}]{
        \includegraphics[width=0.4\columnwidth]{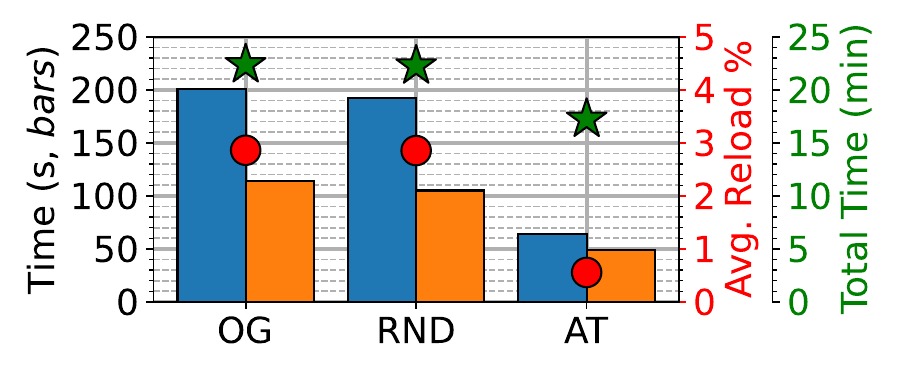}
    }
    \caption{Performance comparison of OG, RND, and AT orderings showing read (reload) and write (eviction) times~(\textit{bars}, left Y axis), avg. \% destination vertices reloaded per chunk~(\textit{red circles}, inner right Y axis), and total execution time~(\textit{green stars}, outer right Y axis).}
    \label{fig:reorder}
\end{figure}

Fig.~\ref{fig:reorder} compares original (OG), random (RND), and ATLAS (AT) orderings for a 2-layer GCN on IL and MA with a 50\,GiB hot store, reporting reload/eviction time, mean reload \%~(i.e., the \% of destination vertices reloaded per chunk), and end-to-end runtime.

AT reduces reload time~(\textit{blue bars}, left Y axis) by $\approx3.3\times$ for IL, from $520$--$530$s to $158$s, and by $\approx3\times$ for MA, from $192$--$200$s to $64$s, compared to OG and RND. The corresponding eviction times~(\textit{orange bars}, left Y axis) decrease by about $2.4\times$ for IL and $2.1$--$2.3\times$ for MA. These reductions translate directly to lower end-to-end execution time (\textit{green stars}, outer right Y axis), dropping from $\approx32$ minutes for IL and $22$ minutes for MA under OG and RND to $20$ minutes and $17$ minutes, respectively, with AT. These reductions are primarily because \at ordering prioritizes vertices that contribute most to completion while limiting fan-out, allowing vertices to finish earlier and reducing the number of partially active states in the hot store. 
We validate this using the \textit{span} of a vertex, defined as the difference between the last and first time it receives a message, which captures how long its state must be maintained. The mean span for AT drops from $\approx39.5$M~(OG/RND) to $11.9$M~($3.3\times$ drop) for IL, and from $\approx42.7$M to $15.2$M~($2.8\times$ drop) for MA, with similar reductions across higher percentiles.

We also see the mean reload \% drop from $\approx7\%$ (OG/RND) to $\approx1.1\%$ with AT on IL, and from $\approx2.9\%$ to $0.5\%$ on MA. This reflects fewer evict--reload cycles under AT due to earlier vertex completion.
Finally, we note that reordering is a one-time cost~($213$s on IL, $258$s on MA) amortized across runs.

\subsection{Impact of \at Eviction Policy}

\begin{figure}[t]
    \centering
    \subfloat[IL/GCN2\label{subfig:eviction-1}]{
        \includegraphics[width=0.4\columnwidth]{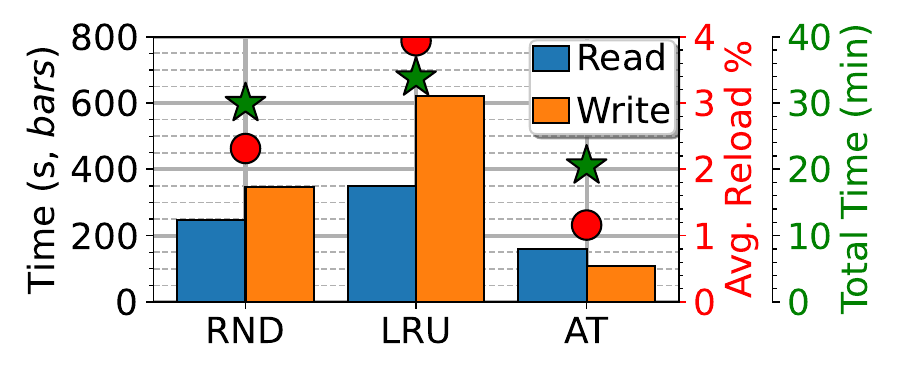}
    }%
    \subfloat[MA/GCN2\label{subfig:eviction-2}]{
        \includegraphics[width=0.4\columnwidth]{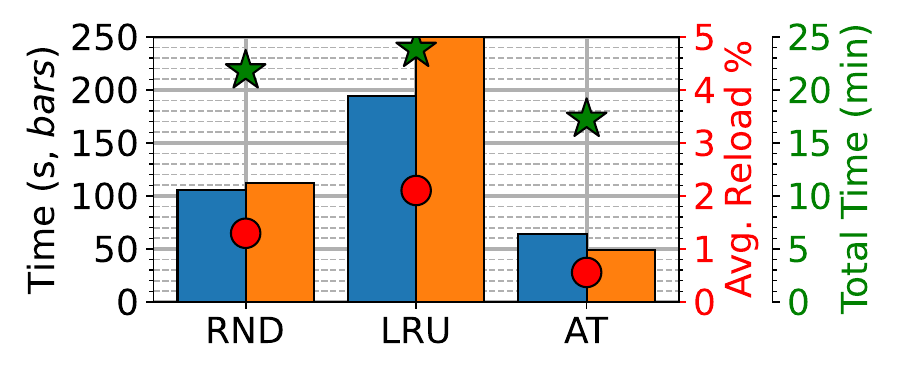}
    }\\
    \subfloat[IL/GCN2\label{subfig:eviction-3}]{
        \includegraphics[width=0.35\columnwidth]{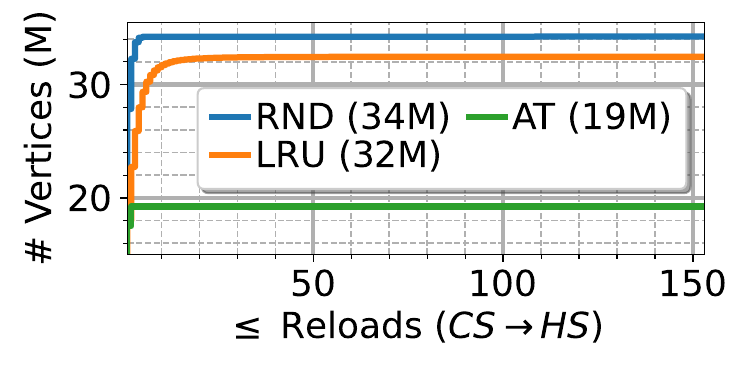}
    }\qquad
    \subfloat[MA/GCN2\label{subfig:eviction-4}]{
        \includegraphics[width=0.35\columnwidth]{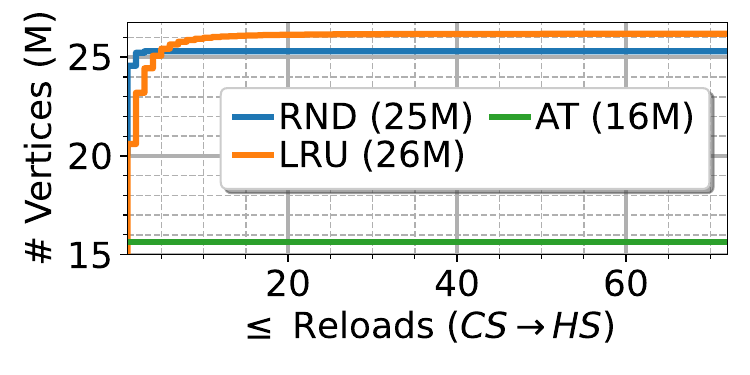}
    }
    \caption{Performance comparison of RND, LRU, and AT eviction policies. \textit{Top Row:} Cold-store read (reload) and write (eviction) times~(\textit{bars}, left Y axis), avg. \% destination vertices reloaded per chunk~(\textit{red circles}, inner right Y axis), and total execution time~(\textit{green stars}, outer right Y axis). 
    \textit{Bottom row:} Cumulative reload distribution of the number of unique vertices reloaded from the cold store.
    }

    \label{fig:eviction}
\end{figure}

Fig.~\ref{fig:eviction} compares AT, RND and LRU eviction for a 2-layer GCN on IL and MA (50\,GiB hot store, AT ordering), reporting cold-store read/write time, mean reload \%, end-to-end runtime, and reload-count distribution.

We notice from Figs.~\ref{subfig:eviction-1} and~\ref{subfig:eviction-2} that AT significantly reduces both reload~(read from SSD) and eviction~(write to SSD) time compared to RND and LRU across both datasets. For IL, AT reduces reload time~(\textit{blue bars}, left Y axis) 
by $1.5$--$2.2\times$ compared to RND/LRU, while eviction time~(\textit{orange bars}, left Y axis) drops by $3$--$5.6\times$. 
Similar trends hold for MA, where AT consistently achieves the lowest I/O cost.
The reduction in read/write times is also reflected in the average reload \%~(\textit{red circles}, inner right Y axis), which drops to approximately $1.2\%$ for IL and $0.6\%$ for MA using AT, compared to about $2.5\%$ and $1.2\%$ for RND, respectively, roughly a $2\times$ reduction for both datasets. This is because AT evicts vertices closest to completion, which significantly reduces the evict-reload cycles.
Notably, LRU performs the worst across both datasets, incurring the highest reload and eviction costs, as it evicts vertices solely by recency, potentially causing high-degree vertices that are still active to be evicted prematurely and repeatedly reloaded as their remaining messages arrive.
This reduction in cold-store I/O directly translates into faster overall execution time~(\textit{green stars}, outer right Y axis), with the total end-to-end execution time decreasing from approximately $30$ minutes with RND to $20$ minutes with AT for IL~($1.5\times$ faster) and from $\approx22$ minutes with RND to $17$ minutes with AT for MA~($1.3\times$ faster).

Fig.~\ref{subfig:eviction-3} and~\ref{subfig:eviction-4} show that AT reduces the number of reloaded vertices by $\approx44\%$ on IL (to $19$M) and $\approx36\%$ on MA (to $16$M), and shortens the reload tail compared to RND/LRU. 
This confirms that minimum-pending-message eviction limits repeated thrashing of partial states.

\subsection{Impact of Hot Store Memory}
\begin{figure}
    \centering
    \subfloat[IL/GCN2\label{subfig:il-mem}]{
        \includegraphics[width=0.4\columnwidth]{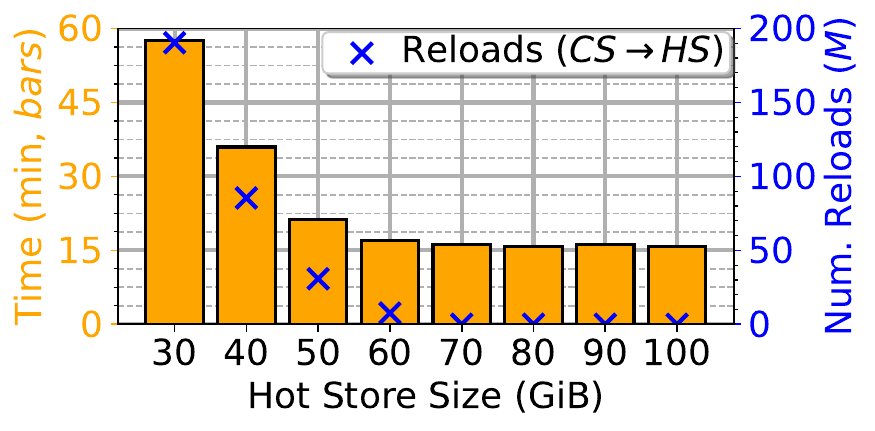}
    }%
    \subfloat[MA/GCN2\label{subfig:ma-mem}]{
        \includegraphics[width=0.4\columnwidth]{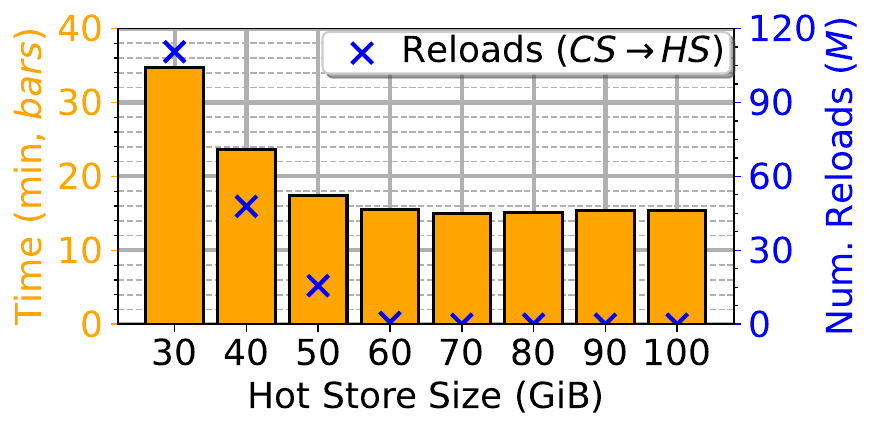}
    }\\
    \subfloat[IL/GCN2/CS\label{subfig:il-mem-cs}]{
        \includegraphics[width=0.35\columnwidth]{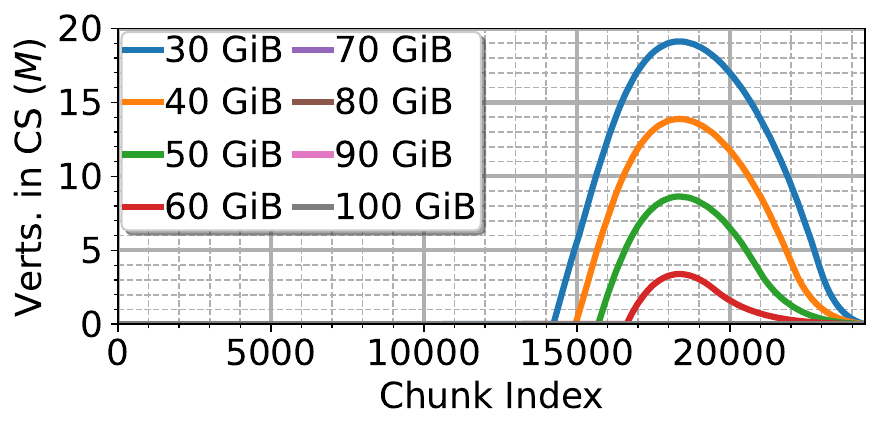}
    }\qquad
    \subfloat[MA/GCN2/CS\label{subfig:ma-mem-cs}]{
        \includegraphics[width=0.35\columnwidth]{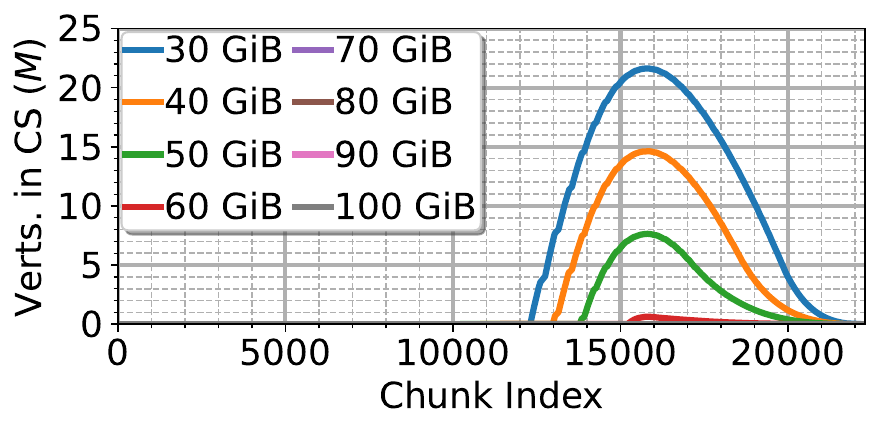}
    }
    \caption{Sensitivity of \at to hot-store memory budgets for IL and MA. \textit{Top Row:}  Execution time~(\textit{bars}, left Y axis) and num. reloads~(\textit{marker}, right Y axis) across memory budgets. 
    \textit{Bottom row:} Num. vertices in \textit{cold store} state across chunks.
    }
    \label{fig:memory}
\end{figure}

Finally, we study the impact of hot-store memory size on the performance of \at for IL and MA datasets using a 2-layer GCN model.
Fig.~\ref{fig:memory} varies hot-store budget (30--100\,GiB) for a 2-layer GCN on IL and MA, reporting end-to-end time and cold-to-hot reloads.
We fix the reordering strategy and eviction policy to AT.

The number of reloads correlates closely with the end-to-end execution time for both IL and MA. As the hot-store size increases, reloads drop sharply, and the runtime decreases accordingly. For IL, increasing the hot-store from $30$\,GiB to $100$\,GiB reduces execution time by $\approx3.8\times$, while for MA, the reduction is $\approx2.3\times$. 
For both datasets, reloads drop to (nearly) zero beyond a memory threshold. For IL, the threshold is 70 GiB; for MA, 60 GiB. Beyond this point, increasing the hot-store size has little to no impact on runtime, suggesting that once the hot store is large enough to avoid evictions, reloads are eliminated and performance stabilizes. In contrast, when the hot store is smaller than this threshold, frequent reloads lead to higher execution time.

We further show the number of vertices in the cold store in Figs.~\ref{subfig:il-mem-cs} and~\ref{subfig:ma-mem-cs} as hot store size increases. The cold-store footprint is large for smaller hot-store sizes~($30$--$50$\,GiB) for both IL and MA, reaching $8$--$22$ million vertices and indicating frequent evictions. As the hot-store size increases, the number of vertices in the cold store drops sharply and becomes negligible beyond 70 GiB.

Overall, \at sustains stable performance on $\approx200$\,GiB feature sets with $\approx60$\,GiB hot-store memory (plus framework overheads).

\section{Related Works}\label{sec:related}
Due to their ability to learn expressive representations from linked data, GNNs have been widely adopted in real-world applications. Social media platforms such as Pinterest use GNNs for content recommendations~\cite {pinsage}, while major e-commerce platforms, including Alibaba, employ GNNs to model user behavior and deliver personalized product recommendations~\cite{aligraph}. Similarly, Google leverages GNNs to estimate travel times in Google Maps~\cite{derrow2021eta}.
In practice, however, these graphs are rarely static. GNNs deployed in such environments are therefore commonly retrained periodically~(e.g., daily or weekly) to address data drift caused by evolving graph structure, vertex features, and label distributions~\cite{rossi2020temporal, xia2023redundancy}. Between retraining cycles, models must continue serving inference requests, requiring access to up-to-date embeddings or prediction results for vertices and edges~\cite{inkstream, ripple}. In this paper, we focus on accelerating the evaluation of GNNs on billion-scale graphs using a single machine, enabling cost-effective inference without distributed infrastructure.

\subsection{Large-scale GNN Training and Inference}
While there are plenty of systems working on scaling GNN \textit{training}~\cite{p3,aligraph,tarafder2025optimization,distdgl}, comparatively little attention has been paid to efficient, large-scale GNN \textit{inference}. Existing industrial systems such as AliGraph and DistDGL are explicitly designed around distributed mini-batch training; these partition graphs across clusters of machines, co-locate data and computation, and optimize sampling and gradient synchronization to train on hundred-million-node graphs~\cite{aligraph,distdgl}. More generally, popular GNN libraries like PyTorch Geometric~\cite{fey2019fast} and DGL~\cite{wang2019deep} provide rich support for mini-batch training and neighborhood sampling on large graphs, but their abstractions and implementations are not optimized for full-graph inference on a single machine. 

Therefore, scaling inference to billion-scale graphs means adopting distributed deployments (e.g., DistDGL-style clusters), which incur substantial setup complexity and infrastructure costs. Distributed full-graph inference is especially challenging because it operates over the entire graph, leading to substantially larger working sets than during training. Additionally, disabling neighborhood sampling for deterministic inference increases inter-node communication due to the need to share embeddings across the cluster, often resulting in out-of-memory~(OOM) failures. 

A few recent works address this challenge by proposing inference-specific frameworks. 
InferTurbo~\cite{inferturbo}, in turn, adopts a GAS-style abstraction with strategies such as \textit{shadow nodes} and \textit{partial-gather} to eliminate redundant k-hop computation and to balance load for hub nodes, enabling full-graph GNN inference on industrial graphs. However, it is also built on a cluster-based infrastructure with MapReduce-based deployments across $5000$ instances.

DGI~\cite{dgi} translates existing training code into a \textit{layer-wise} execution plan and supports out-of-core full-graph by dynamically batching nodes and reordering them to improve input sharing, achieving substantial speedups over naive layer-wise baselines on large graphs. While DGI demonstrates strong performance for in-memory inference on a single machine, its reliance on a memory-mapped out-of-core implementation limits scalability when graphs exceed available RAM, resulting in high read overheads. We compare \at against DGI in our experiments.

\subsection{Out-of-core GNN Training Systems}
Owing to the infrastructure costs and the complexity of implementing distributed GNN training, many recent works have explored ways to utilize all available resources, particularly disk, on an isolated commodity machine for GNN training on billion-scale graphs~\cite{capsule, outre, marius, diskgnn, ginex, hyperion, caliex}. Most of these works mainly focus on optimizing data transfer, organization, and layout to speed up training. DiskGNN~\cite{diskgnn} introduces an offline sampling paradigm that decouples graph sampling from model computation, precomputes many mini-batch samples, and then packs node features contiguously on disk to avoid read amplification. To further reduce disk traffic, it uses a four-level feature store, batched feature packing, and a pipelined training pipeline. However, it must generate all computation graphs beforehand, which is a costly step when performing full-graph GNN inference over the entire graph. Additionally, DiskGNN relies heavily on feature packing for each computation graph as a contiguous chunk on disk. While this is a good strategy for training, for full-graph inference it leads to a massive blowup in disk space, since each computation involves many more vertices. 

Capsule~\cite{capsule} combines graph partitioning with subgraph pruning so that each training subgraph fits the GPU budget and designs a subgraph loading mechanism modeled as a shortest Hamiltonian cycle to minimize the cost of loading successive subgraphs from disk. Inspired by the inspector-executor model from compilers, Ginex~\cite{ginex} splits training into a sample stage and a gather stage, allowing it to know in advance which node features future mini-batches will require. Ginex applies Belady's optimal caching algorithm to maintain feature vectors in memory. However, full-batch inference accesses span the entire graph, leading to cache thrashing and massive random SSD reads. 

\section{Conclusion}
\label{sec:conclude}
In this paper, we have proposed \at, a broadcast-driven, OOC framework that enables efficient full-batch GNN inference on billion-scale graphs that do not fit in memory. By restructuring inference around sequential, single-pass I/O and introducing a tiered memory-disk hierarchy with an intelligent eviction policy, \at overcomes the challenges of read amplification, memory pressure, and irregular access patterns inherent in traditional gather-based methods. Our end-to-end pipelined design further overlaps disk access, message aggregation, and GPU-based inference to sustain high throughput under modest memory budgets available on a single workstation. Experiments on billion-sized graphs taking up to $550$\,GiB on disk demonstrate substantial I/O reduction and up to $12$--$30\times$ speedups over SOTA baselines, several of which fail to complete. 

\at offers a cost-effective design for large-scale GNN inference on single-machine platforms and opens opportunities to extend this approach to attention-based layers and incremental inference workloads, as well as to other GNN-specific additions, such as neighborhood-sampling-based inference.

\clearpage
\bibliographystyle{plain}
\bibliography{ref}
\clearpage
\end{document}